\documentclass{JHEP3}

\usepackage{amsmath,epsfig,bbm}

\newcommand{\tmop}[1]{\operatorname{#1}}
\newcommand{\mathi}{\mathrm{i}}
\newcommand{\emdash}{---}
\newcommand{\tmem}[1]{{\em #1\/}}
\newcommand{\tmfloatcontents}{}
\newlength{\tmfloatwidth}
\newcommand{\tmfloat}[5]{
  \renewcommand{\tmfloatcontents}{#4}
  \setlength{\tmfloatwidth}{\widthof{\tmfloatcontents}+1in}
  \ifthenelse{\equal{#2}{small}}
    {\ifthenelse{\lengthtest{\tmfloatwidth > \linewidth}}
      {\setlength{\tmfloatwidth}{\linewidth}}{}}
    {\setlength{\tmfloatwidth}{\linewidth}}  \begin{minipage}[#1]{\tmfloatwidth}
    \begin{center}
      \tmfloatcontents
      \captionof{#3}{#5}
    \end{center}
  \end{minipage}}
\newcommand{\mathe}{\mathrm{e}}
\newcommand{\mathd}{\mathrm{d}}
\newcommand{\mathpi}{\pi}

\catcode`\à=\active \defà{\`a} \catcode`\À=\active \defÀ{\`A}
\catcode`\á=\active \defá{\'a} \catcode`\Á=\active \defÁ{\'A}
\catcode`\ä=\active \defä{\"a} \catcode`\Ä=\active \defÄ{\"A}
\catcode`\â=\active \defâ{\^a} \catcode`\Â=\active \defÂ{\^A}
\catcode`\å=\active \defå{{\aa}} \catcode`\Å=\active \defÅ{{\AA}}
\catcode`\ç=\active \defç{\c{c}} \catcode`\Ç=\active \defÇ{\c{C}}
\catcode`\è=\active \defè{\`e} \catcode`\È=\active \defÈ{\`E}
\catcode`\é=\active \defé{\'e} \catcode`\É=\active \defÉ{\'E}
\catcode`\ë=\active \defë{\"e} \catcode`\Ë=\active \defË{\"E}
\catcode`\ê=\active \defê{\^e} \catcode`\Ê=\active \defÊ{\^E}
\catcode`\ì=\active \defì{\`{\i}} \catcode`\Ì=\active \defÌ{\`{\I}}
\catcode`\í=\active \defí{\'{\i}} \catcode`\Í=\active \defÍ{\'{\I}}
\catcode`\ï=\active \defï{\"{\i}} \catcode`\Ï=\active \defÏ{\"{\I}}
\catcode`\î=\active \defî{\^{\i}} \catcode`\Î=\active \defÎ{\^{\I}}
\catcode`\ò=\active \defò{\`o} \catcode`\Ò=\active \defÒ{\`O}
\catcode`\ó=\active \defó{\'o} \catcode`\Ó=\active \defÓ{\'O}
\catcode`\ö=\active \defö{\"o} \catcode`\Ö=\active \defÖ{\"O}
\catcode`\ô=\active \defô{\^o} \catcode`\Ô=\active \defÔ{\^O}
\catcode`\ù=\active \defù{\`u} \catcode`\Ù=\active \defÙ{\`U}
\catcode`\ú=\active \defú{\'u} \catcode`\Ú=\active \defÚ{\'U}
\catcode`\ü=\active \defü{\"u} \catcode`\Ü=\active \defÜ{\"U}
\catcode`\û=\active \defû{\^u} \catcode`\Û=\active \defÛ{\^U}
\catcode`\ý=\active \defý{\'y} \catcode`\Ý=\active \defÝ{\'Y}
\catcode`\ÿ=\active \defÿ{\"y} \catcode`\˜=\active \def˜{\"Y}
\catcode`\½=\active \def½{!`}
\catcode`\¾=\active \def¾{?`}
\catcode`\ß=\active \defß{{\ss}}

\title{Anomaly-matching and Higgs-less effective theories} 
\author{Johannes Hirn$^{a, b}$ and Jan Stern$^{a}$\\
\\
$^a$ Groupe Physique Théorique\thanks{\em{Unit\'e mixte de recherche 8608.}}, IPN Orsay,\\
$^{\ }$ Université~Paris-Sud~XI, 91406~Orsay, France\\ 
\\
$^b$ Institute for Particle Physics Phenomenology,\\
$^{\ }$ University of  Durham, Durham~DH1~3LE, UK\\
\\
$^{\ }$ E-mails: \email{johannes.hirn@durham.ac.uk}, \email{stern@ipno.in2p3.fr}}

\abstract{We reconsider the low-energy effective theory for Higgs-less electroweak
symmetry breaking: we study the anomaly-matching in the
situation where all Goldstone fields disappear from the spectrum as a result
of the Higgs mechanism. We find that the global $\tmop{SU}\left(2\right)_L \times \tmop{SU}\left(2\right)_R \times \tmop{U}\left(1\right)_{B-L}$ symmetry of the underlying theory, which is spontaneously broken to $\tmop{SU}(2)_{L+R}\times \tmop{U}(1)_{B-L}$, has to be anomaly-free.
For the sake of
generality, we include the possibility of light spin-$1/2$
bound states resulting from the dynamics of the strongly-interacting
symmetry-breaking sector, in addition to the Goldstone bosons. Such composite
fermions may have non-standard couplings at the leading order, and an
arbitrary total~$B-L$ charge. In order to perform the
anomaly-matching in that case, we generalize the construction of the
Wess-Zumino effective lagrangian.
Composite fermions beyond the three
known generations are theoretically allowed, and there are no restrictions
from the anomaly-matching on their couplings nor on their~$\mathrm{U}\left(1\right)_{B-L}$
charge.
Absence of global anomalies for the composite sector as a whole does not preclude anomalous triple gauge boson couplings arising from
composite fermion triangular diagrams.
On the other hand, the trace of~$B-L$ over elementary
fermions must vanish if all Goldstone modes are to disappear from the
spectrum.}


\preprint{\texttt{IPNO-DR-0402}, \texttt{IPPP/04/10}, \texttt{DCPT/04/20}}

\begin{document}

\section{Introduction}

The detailed dynamics of the electroweak symmetry breaking (EWSB)  is so far unknown, and the possibility that it may be strongly-interacting above a few TeV scale is not ruled out.
Below that scale, a systematic 
description would then be provided by a Low-Energy Effective Theory
(LEET) following the lines initiated by Weinberg~{\cite{Weinberg:1979kz}}. In the limit of 
small momenta the effective theory remains weakly coupled and the 
relevant degrees of freedom remain light due to the underlying chiral and 
gauge symmetries. The LEET is formulated as a systematic expansion in 
powers of momenta. The list of light fields relevant at low energies 
whose mass is protected by a symmetry is then a required input. Here,
we will assume a minimal content of the composite EWSB sector:
the only light bosonic
fields are the three Goldstone modes arising from the spontaneous breakdown of the symmetry~$\tmop{SU}\left(2\right)_L \times \tmop{SU}\left(2\right)_R \times \mathrm{U}\left(1\right) \longrightarrow \tmop{SU}\left(2\right)_{L+R} \times \mathrm{U}\left(1\right)$, 
which is supposed to characterize the 
strongly-interacting symmetry-breaking sector at low energies  (the underlying theory will be referred to as `techni-theory').
In addition we introduce elementary fields external to the techni-theory: $\tmop{SU}\left(2\right)\times \tmop{SU}\left(2\right)$ Yang-Mills fields and the elementary fermion doublets.
Once the elementary sector and the composite sectors are coupled, the three Goldstone bosons produce the masses of the $W^{\pm}$ and $Z^0$ in the usual way.
This scenario corresponds to the case of Higgs-less symmetry breaking, in which no light scalar remains in the spectrum.

Renormalization is then performed order-by-order in the low-energy expansion.
Similarly, unitarity is gradually restored in the same momentum expansion, rather than in
powers of coupling constants as would be the case for a renormalizable theory.
In a LEET, even though observables do not depend on a cut-off or on a renormalization scale, 
there is an energy scale inherent to the theory  above which the low-energy expansion becomes
unreliable. In the case of Chiral Perturbation Theory ($\chi$PT)~{\cite{Gasser:1984yg,Gasser:1985gg}}, this scale is known to be given by $\Lambda \simeq 4
\mathpi f$ {\cite{Georgi:1985kw}}, where $f$ is the pion decay constant, although one may hope that the inclusion of additional states
(resonances) in the LEET can push this scale up.  In the case of EWSB, a similar reasoning leads to an estimate of~$\Lambda \simeq 3\ \text{TeV}$ .

At a given level of accuracy,~$\mathcal{O}\left(p^2\right)$, $\mathcal{O}\left(p^4\right)$,..., new renormalized low-energy constants appear in addition to loop contributions. Precision experiments can then be used to fix these constants, and this has been partially done previously, see e.g.~{\cite{Dobado:1991zh,Nyffeler:1999hp}}. 
The values of these constants reflect the details of the unknown dynamics of the techni-theory.
In the past, different models for the techni-theory and for Higgs-less EWSB have been considered:
for instance technicolor models~{\cite{Susskind:1979ms}}, directly inspired by QCD, suggest values of~$\mathcal{O}\left(p^4\right)$ low-energy constants, which are at variance with electroweak precision tests~{\cite{Holdom:1990tc,GOlden:1991ig,Peskin:1992sw,Hagiwara:2002fs,Hill:2002ap}}.  For a more recent discussion, see also~\cite{Nyffeler:1999ap}.
More refined versions such as walking technicolor~{\cite{Holdom:1981rm,Holdom:1985sk,Yamawaki:1986zg,Appelquist:1986an,Appelquist:1987tr}}, are not yet completely excluded~{\cite{Hagiwara:2002fs,Hill:2002ap}}.
Another particular example is the heavy Higgs limit of the Standard Model~(SM)~{\cite{Appelquist:1980vg,Longhitano:1980iz,Longhitano:1981tm,Nyffeler:1999ap}}.
More recently, Higgs-less models~{\cite{Csaki:2003zu,Barbieri:2003pr,Cacciapaglia:2004jz}} have been
revived, in connection with five-dimensional set-ups.
The variety of this (incomplete) list incites us to approach the question in a model-independent way within a LEET, trying to understand the generic features of Higgs-less EWSB.

In the Higgs-less LEET, difficulties arise first at the leading~$\mathcal{O} \left( p^2 \right)$ order: 
given the standard power counting and the~$\tmop{SU}\left(2\right) \times \tmop{U}\left(1\right)$ symmetry of the problem, one finds~$\mathcal{O} \left( p^2 \right)$ operators, which are not observed experimentally.
The problem is then to suppress these unwanted terms {\em naturally} ---i.e. in connection with a symmetry.
A systematic framework for the resolution of this problem has been
proposed in~{\cite{Hirn:2004ze}}:
one appeals to a larger symmetry, characteristic of the situation where the elementary gauge and elementary
fermion sector on one hand and the composite symmetry-breaking sector on the other hand are decoupled. 
This larger natural symmetry is~$\left[\tmop{SU}\left(2\right)\times\tmop{SU}\left(2\right)\right]^2\times \tmop{U}\left(1\right)$.
The link between the two sectors is introduced via non-propagating spurion fields, which are assumed to be covariantly constant.
The solution of these constraints is invariant under a reduced symmetry~$\tmop{SU} \left( 2 \right) \times \mathrm{U} \left(1 \right)$, introducing the coupling between composite Goldstone modes and elementary fields.
Using the relics of the spurions as small expansion parameters,
the scheme gives a consistent description for the suppression of the
aforementioned unwanted terms, namely non-standard couplings of fermions to vector
bosons, and tree-level~$S$ parameter.
In addition, it allows for a description
of fermion masses including mass-splittings within doublets at the same order
as the masses themselves, standard CKM mixing and $C P$~violation.

In this paper, we address a second potential problem of Higgs-less effective theories concerning anomaly-matching~\cite{'tHooft:1979bh}, which requires clarification.
The anomalous Ward identities for the symmetry currents of the techni-theory have to be reproduced in the LEET by light spin-$0$ or spin-$1/2$ physical states~{\cite{Coleman:1982yg}}.
In the absence of the latter, this amounts to the construction of a Wess-Zumino
effective lagrangian involving the Goldstone fields~{\cite{Wess:1971yu}}.
It is not a priori clear how the anomaly-matching proceeds if all spin-$0$ physical states disappear from the spectrum due to the Higgs mechanism.
Therefore, we consider the possibility of
light composite fermions as bound states of the techni-theory and investigate
the anomaly-matching condition in a Higgs-less theory with this more general setting. Note that the distinction between composite and elementary fermions from the
point of view of the underlying theory is clear: the first ones are not
fundamental variables, but appear among the bound states as a consequence
of the strong dynamics of the techni-theory, while the latter are introduced
as fundamental fields external to the techni-theory. Therefore, elementary
fermions do not participate in the anomaly-matching, given that they are
described by the same fields both in the low and high-energy descriptions.

The paper is organized as follows.

In section~\ref{sec: sym}, we give an overview of the content of the composite and elementary sectors, and their coupling via spurions.
We show that, contrary to elementary fermions, composite fermions
have non-standard couplings at~$\mathcal{O} \left( p^2 \right)$.

Section \ref{s:anomalies} deals with the
anomaly-matching between the techni-theory and its effective theory in the
general case: we first study the variation of the generating functional for
the Noether currents of the techni-theory. This variation is reproduced by a
standard Wess-Zumino lagrangian in the absence of composite fermions. We show
that, if such fermions are present, it is always possible to build a generalized
Wess-Zumino action  such that the sum of its variation and the variation of the
composite fermion determinant  reproduces the anomalies of the techni-theory, whatever the values of the non-standard couplings of the composite
fermions.
Therefore, the couplings of these composite fermions are unconstrained,
including the value of their $\mathrm{U} \left( 1 \right)_{B - L}$ charge:
there is no such constraint as the vanishing of the trace of $B-L$ over composite
fermions.

The main result of the anomaly-matching in Higgs-less theories is presented in section~\ref{s:U-gauge}.
It concerns the techni-theory itself:
its low-energy symmetry~$\tmop{SU}\left(2\right)_L \times \tmop{SU}\left(2\right)_R \times \mathrm{U}\left(1\right)_{B-L}$ must be anomaly-free if all Goldstone modes are to disappear
from the spectrum via the Higgs mechanism.
This statement is independent of the presence of composite fermions.
The vanishing of the trace of~$B - L$ for elementary fermions
is still required in order to allow for a consistent formulation of the Higgs-less EWSB.
Notice that the logical status of this result is quite different from what
happens in the SM, where it stems from the requirement of renormalizability.

Finally, in section \ref{s:triangles}, we give the appropriate formulation for the theory in the unitary gauge.
In particular, we study the contribution of triangular diagrams involving composite fermions to triple gauge boson vertices.

We conclude in section \ref{s:concl} by summarizing our main results and
mentioning open questions.

\section{Composite and elementary sectors} \label{sec: sym}

In this section we describe the LEET at lowest order, distinguishing between
the composite sector resulting from the dynamics of the techni-theory and the
elementary sector. In the bosonic sector, only the four elementary gauge
fields that become the $W^{\pm}$, $Z^0$ and  photon are introduced, together
with the three real Goldstone bosons (composite fields) required to give them
masses. These Goldstone bosons describe the spontaneous breaking of symmetry
$\tmop{SU} \left( 2 \right)_L \times \tmop{SU} \left( 2 \right)_R
\longrightarrow \tmop{SU} \left( 2 \right)_{L + R}$ assumed to occur as a
result of the strong dynamics of the techni-theory.
In addition, we consider the case where the theory has a global vector symmetry $\tmop{U}\left(1\right)_{B - L}$, which is required for our construction \footnote{In fact, the identification of the $B-L$ transformation with the third component of the right isospin (to obtain the hypercharge transformation) will be obtained via the constraint applied on the spurion $\phi$. This procedure allows us to maintain the $\tmop{SU}\left(2\right)$ custodial symmetry in the lagrangian.}.

The generating functional for the seven Noether currents~$J_L^{a
\mu}, J_R^{a \mu}$, for~$a=1,2,3$, and~$J_{B - L}^{\mu}$ that correspond to global symmetries
of the techni-theory can be formally defined through the relation
\begin{eqnarray}
  \mathe^{\mathi \Gamma \left[ L_{\mu}, R_{\mu}, B^0_{\mu} \right]} & = & \int
  \mathd \left[ T \right] \mathe^{\mathi \int \mathd x\mathcal{L}_{\tmop{TT}}
  \left[ T \right] + J_L^{a \mu} \left[ T \right] L_{\mu}^a + J_R^{a \mu}
  \left[ T \right] R_{\mu}^a + J_{B - L}^{\mu} \left[ T \right] B^0_{\mu}} . 
  \label{aa}
\end{eqnarray}
Here $T$ collectively denotes the fundamental fields of the techni-theory and
$\mathcal{L}_{\tmop{TT}}$ is its {\emdash}unspecified{\emdash} lagrangian. The
sources $L^a_{\mu}, R^a_{\mu}, B^0_{\mu}$ in principle allow us to extract
Green's functions for the corresponding Noether currents. In fact, the reason
we resort to a LEET stems from our ignorance: we do not know what the
appropriate variables $T$ are. Instead, we work with the representation of the techni-theory in
terms of the low-energy variables, defined as an expansion in powers of
momenta and other appropriate parameters~{\cite{Weinberg:1979kz}}. This
expansion relies on the symmetries of the theory and the fact that the
currents have to satisfy both normal and anomalous Ward identities. The latter
are known up to a multiplicative constant, which reflects the field content of
the techni-theory at higher energies. The anomalous Ward identities obtained
from the fundamental theory have to be reproduced by the LEET itself. Before
we turn to this anomaly-matching, we describe the low-energy representation of
the symmetry-breaking sector, and then the elementary sector, which is external
to the techni-theory. We also briefly review how the coupling between the two
sectors is introduced via spurions.

\subsection{Low-energy representation of the symmetry-breaking sector}
\label{s:intr-comp}

\EPSFIGURE{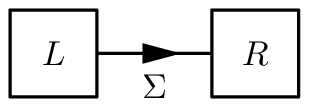, width=.25\textwidth}{Transformation properties of the Goldstone bosons.\label{f:Sigma}}

The minimal field content of the symmetry-breaking sector is a triplet of
Goldstone bosons. As schematized in figure \ref{f:Sigma}, the Goldstone modes
are parameterized in the LEET by an $\tmop{SU} \left( 2 \right)$ matrix
$\Sigma$ transforming as
\begin{eqnarray}
  \Sigma & \longmapsto & L \Sigma R^{\dag},  \label{00000}
\end{eqnarray}
with $\left( L, R \right) \in \tmop{SU} \left( 2 \right)_L \times \tmop{SU}
\left( 2 \right)_R$. In order to deal with the Noether currents generated by
the symmetries of the techni-theory, we consider local $\tmop{SU} \left( 2
\right)_L \times \tmop{SU} \left( 2 \right)_R$ transformations and introduce
the same sources first used in (\ref{aa}) as connections in the covariant
derivative~{\cite{Gasser:1984yg}}
\begin{eqnarray}
  D_{\mu} \Sigma & = & \partial_{\mu} \Sigma - \mathi L_{\mu} \Sigma + \mathi
  \Sigma R_{\mu} .  \label{00100}
\end{eqnarray}
Up to anomalies, the lagrangian must be invariant if we simultaneously perform
the following transformations on the sources {\cite{Leutwyler:1994iq}}
\begin{eqnarray}
  L_{\mu} & \longmapsto & LL_{\mu} L^{\dag} + \mathi L \partial_{\mu}
  L^{\dag}, \\
  R_{\mu} & \longmapsto & RR_{\mu} R^{\dag} + \mathi R \partial_{\mu} R^{\dag}
  . 
\end{eqnarray}
The power counting for the low-energy expansion is known
{\cite{Gasser:1984yg}}
\begin{eqnarray}
  \partial_{\mu}, L_{\mu}, R_{\mu}, B_{\mu}^0 & = & \mathcal{O} \left( p^1
  \right),  \label{pow}\\
  f, \Sigma & = & \mathcal{O} \left( p^0 \right), 
\end{eqnarray}
where $f$ is the Goldstone boson decay constant. The energy scale $\Lambda$
to which the momenta have to be compared to is known to be approximately
equal to $4 \pi f$ {\cite{Georgi:1985kw}}. The leading-order $\mathcal{O}
\left( p^2 \right)$ lagrangian is then
\begin{eqnarray}
  \mathcal{L}_{\tmop{SB}} & = & \frac{f^2}{4}  \left\langle D_{\mu} \Sigma
  D^{\mu} \Sigma^{\dag} \right\rangle, 
\end{eqnarray}
where $\left\langle \cdots \right\rangle$ denotes the trace of a two-by-two
matrix. The full effective lagrangian is to be constructed as the most general
expression satisfying the symmetry requirements, organized as an
order-by-order expansion in powers of momenta.

It is an open possibility, though not a necessity, that the techni-theory also
produces light spin-$1 / 2$ bound states: composite fermions. These would then
transform under the same  $\tmop{SU} \left( 2 \right)_L \times \tmop{SU}
\left( 2 \right)_R$ transformation as the Goldstone bosons, which are the
other bound states of the techni-theory. By including composite fermions in
the LEET, we are implicitly assuming that they are light with respect to the
scale $\Lambda$. This can be understood for instance if they play a role in
the anomaly-matching {\cite{'tHooft:1979bh}}. Consequently, light composite
fermions have to be included in a discussion of anomaly-matching for the sake
of completeness.

The left and right-handed composite fermion doublets are
\begin{eqnarray}
  \chi^{\text{c}}_{L, R} & = & \frac{1 \mp \gamma_5}{2} \chi^{\text{c}}, 
\end{eqnarray}
with the appropriate counting given by {\cite{Wudka:1994ny,Nyffeler:1999ap}}
\begin{eqnarray}
  \chi^{\text{c}} & = & \mathcal{O} \left( p^{1 / 2} \right) . 
\end{eqnarray}
These fermions are taken to have identical charges under the $\mathrm{U}
\left( 1 \right)_{B - L}$ symmetry {\footnote{In order to conform to the
standard notation dealing with anomalies, we will call $\alpha^0$ the gauge
function instead of $\beta^0$ used in {\cite{Hirn:2004ze}}.}}
\begin{eqnarray}
  \chi^{\text{c}}_L & \longmapsto & \text{}^t \chi^{\text{c}}_L = R \mathe^{- \mathi \frac{B - L}{2}
  \alpha^0} \chi^{\text{c}}_L,  \label{00200}\\
  \chi^{\text{c}}_R & \longmapsto & \text{}^t \chi^{\text{c}}_R = L \mathe^{- \mathi \frac{B - L}{2}
  \alpha^0} \chi^{\text{c}}_R .  \label{00210}
\end{eqnarray}
Here, $B - L$ is just a generic parameter characterizing each composite
fermion doublet.

\EPSFIGURE{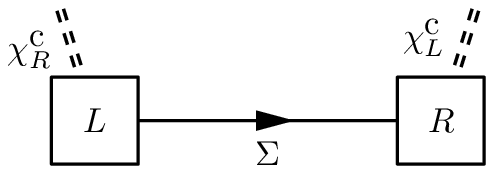}{\label{fig:fermions}Transformation
properties of the composite fermions.}

Except for the $\mathrm{U} \left( 1 \right)_{B - L}$ symmetry, the
transformation properties are summarized in Figure \ref{fig:fermions}. We will
mention briefly at the end of section \ref{sec: matching} why other
possibilities for the transformation properties of these fermions are not of
interest in the context of this paper, in connection with the
anomaly-matching. The most general lagrangian of  order $\mathcal{O} \left(
p^2 \right)$ involving composite fermions and invariant under $\tmop{SU}
\left( 2 \right)_L \times \tmop{SU} \left( 2 \right)_R \times \mathrm{U}
\left( 1 \right)_{B - L}$ is
\begin{eqnarray}
  \mathcal{L}_{\tmop{comp}. \tmop{ferm}.} & = & \mathi
  \overline{\chi^{\text{c}}} \gamma^{\mu} D_{\mu} \chi^{\text{c}} + \mathi
  \delta_L  \overline{\chi^{\text{c}}_L}  \left( \Sigma^{\dag} D_{\mu} \Sigma
  \right) \chi_L^{\text{c}} + \mathi \delta_R  \overline{\chi^{\text{c}}_R} 
  \left( \Sigma D_{\mu} \Sigma^{\dag} \right) \chi_R^{\text{c}} \nonumber\\
  & + & \text{four-fermion interactions},  \label{ar}
\end{eqnarray}
where the four-fermion interactions are suppressed by an unknown scale, which
has no obvious relation to the scale $\Lambda$ introduced above in connection
with Goldstone fields. Consequently, we will neglect these four-fermion
interactions in the sequel. The covariant derivative in (\ref{ar}) is given by
\begin{eqnarray}
  D_{\mu} \chi^{\text{c}} & = & \partial_{\mu} \chi^{\text{c}} - \mathi
  \left\{ \frac{B - L}{2} B^0_{\mu} + \frac{1 - \gamma_5}{2} R_{\mu} + \frac{1
  + \gamma_5}{2} L_{\mu} \right\} \chi^{\text{c}},  \label{Dcomp}
\end{eqnarray}
where the $\mathrm{}$ $\mathrm{U} \left( 1 \right)_{B - L}$ connection
occurring in this covariant derivative transforms as follows
\begin{eqnarray}
  B^0_{\mu} & \longmapsto & B^0_{\mu} - \partial_{\mu} \alpha^0 . 
\end{eqnarray}
Since composite fermions are directly coupled to the Goldstone bosons, a gauge
invariant mass term can be included in $\mathcal{L}_{\tmop{comp}. \tmop{ferm}
.}$ (\ref{ar})
\begin{eqnarray}
  \mathcal{L}_{\tmop{comp}. \tmop{ferm}.} & \longmapsto &
  \mathcal{L}_{\tmop{comp}. \tmop{ferm}.} - m \left(
  \overline{\chi^{\text{c}}_L} \Sigma^{\dag} \chi^{\text{c}}_R +
  \overline{\chi^{\text{c}}_R} \Sigma \chi^{\text{c}}_L \right) .  \label{m}
\end{eqnarray}
There is nothing, in what we have described up to now, that forces the
coefficient $m$ appearing in front of this operator to be small. Notice however that, in order
for this last term in (\ref{m}) to be counted as $\mathcal{O} \left( p^2
\right)$ together with the kinetic term in (\ref{ar}) so as to reproduce the
pole in the propagator, $m$ should be counted as $\mathcal{O} \left( p^1
\right)$. This is related to the need for a mechanism to keep this mass light:
we are assuming that such a mechanism is at play, and therefore use the
counting
\begin{eqnarray}
  m & = & \mathcal{O} \left( p^1 \right) . 
\end{eqnarray}
At this level, mass-splittings within doublets are forbidden by the symmetry
$\tmop{SU} \left( 2 \right)_L \times \tmop{SU} \left( 2 \right)_R \times
\mathrm{U} \left( 1 \right)_{B - L}$.

We have found that non-standard couplings $\delta_L$, $\delta_R$, whose strength are not fixed
{\cite{Appelquist:1985rr,Peccei:1990kr}}, are already present at
leading-order in (\ref{ar}), due to the symmetry being realized non-linearly
on the Goldstone fields {\footnote{In the standard elementary Higgs model,
these non-standard terms would appear as dimension six operator.}}. For
elementary fermions, such non-standard couplings are suppressed since they
necessarily involve spurions, see {\cite{Hirn:2004ze}}.

This completes for the time being the effective low-energy description of the
composite sector: the low-energy representation for the generating functional
(\ref{aa}) is then
\begin{eqnarray}
 &&  \mathe^{\mathi \Gamma \left[ L_{\mu}, R_{\mu}, B^0_{\mu} \right]}\nonumber\\
 & = & \int
  \mathd \left[ \Sigma \right] \mathd \left[ \chi^{\text{c}} \right]
  \mathe^{\mathi \int \mathd x\mathcal{L}_{\tmop{SB}} \left[ \Sigma, L_{\mu},
  R_{\mu}, B^0_{\mu} \right] +\mathcal{L}_{\tmop{comp}. \tmop{ferm}.} \left[
  \chi^{\text{c}}, \Sigma, L_{\mu}, R_{\mu}, B^0_{\mu} \right] +\mathcal{O}
  \left( p^4 \right)} .  \label{rep}
\end{eqnarray}
We now turn to the elementary sector, and to the coupling between the two
sectors via spurions.

\subsection{The elementary sector and spurion-induced couplings}
\label{s:spurions}

The elementary sector is external to the techni-theory: in the low-energy
description, this results in a separation between the composite sector
described in section \ref{s:intr-comp}, and the elementary sector. The two
sectors will only be coupled via spurions.

Among the elementary fields are the gauge fields, which we introduce as an
$\tmop{SU} \left( 2 \right)_{G_0} \times \tmop{SU} \left( 2 \right)_{G_1}$
Yang-Mills theory, with the connections transforming as
\begin{eqnarray}
  G_{0 \mu} & \longmapsto & G_0 G_{0 \mu} G_0^{\dag} + \frac{\mathi}{g_0} G_0
  \partial_{\mu} G_0^{\dag}, \\
  G_{1 \mu} & \longmapsto & G_1 G_{1 \mu} G_1^{\dag} + \frac{\mathi}{g_1} G_1
  \partial_{\mu} G_1^{\dag}, 
\end{eqnarray}
where $\left( G_0, G_1 \right) \in \tmop{SU} \left( 2 \right)^2$. With the
usual power counting {\cite{Urech:1995hd,Wudka:1994ny}}
\begin{eqnarray}
  G_{0 \mu}, G_{1 \mu} & = & \mathcal{O} \left( p^0  \right), \\
  g_0, g_1 & = & \mathcal{O} \left( p^1 \right), 
\end{eqnarray}
the $\mathcal{O} \left( p^2 \right)$ lagrangian is then simply
\begin{eqnarray}
  \mathcal{L}_{\text{gauge}} & = & - \frac{1}{2}  \left\langle G_{1 \mu \nu}
  G_1^{\mu \nu} \right\rangle - \frac{1}{2}  \left\langle G_{0 \mu \nu}
  G_0^{\mu \nu} \right\rangle . 
\end{eqnarray}
In the above, the field-strengths are as usual
\begin{eqnarray}
  G_{k \mu \nu} & = & \partial_{\mu} G_{k \mu} - \partial_{\nu} G_{k \mu} -
  \mathi g_k  \left[ G_{k \mu}, G_{k \nu} \right], \hspace{2em} \tmop{for}
   k = 0, 1 . 
\end{eqnarray}
Elementary fermions doublets $\chi^{\text{e}}$ are directly coupled to the
gauge fields of $\tmop{SU} \left( 2 \right)_{G_0} \times \tmop{SU} \left( 2
\right)_{G_1}$, rather than being charged under the symmetries of the
techni-theory $\tmop{SU} \left( 2 \right)_L \times \tmop{SU} \left( 2
\right)_R$ as is the case for composite fermions $\chi^c$ (see section
\ref{s:intr-comp}). The elementary fermions transform as
\begin{eqnarray}
  \chi^{\text{e}}_L & \longmapsto &  G_1
  \mathe^{- \mathi \frac{B - L}{2} \alpha^0} \chi^{\text{e}}_L,  \label{cp}\\
  \chi^{\text{e}}_R & \longmapsto & G_0
  \mathe^{- \mathi \frac{B - L}{2} \alpha^0} \chi^{\text{e}}_R .  \label{cq}
\end{eqnarray}
\EPSFIGURE{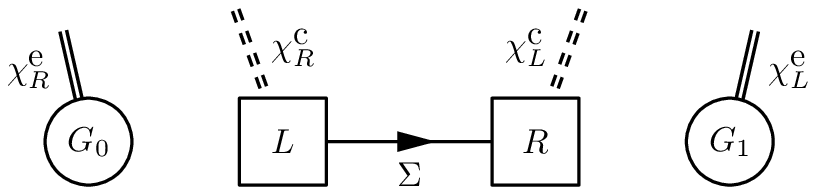, width=.35\textwidth}{The composite and elementary sectors decoupled.\label{f:elem}}

Note that the value of $B - L$ for elementary fermion and composite fermion
doublets are a priori different. The complete $\mathcal{O} \left( p^2 
\right)$ lagrangian describing elementary fermions reads
\begin{eqnarray}
  \mathcal{L}_{\tmop{elem}. \tmop{ferm}.} & = & \mathi \overline{\chi^{\text{e}}}
  \gamma^{\mu} D_{\mu} \chi^{\text{e}} \nonumber\\
&+& \text{four-fermion interactions}, 
  \label{cr}
\end{eqnarray}
where
\begin{eqnarray}
  D_{\mu} \chi^{\text{e}} & = & \partial_{\mu} \chi^{\text{e}} - \mathi
  \left\{ \frac{B - L}{2} B^0_{\mu} + \frac{1 - \gamma_5}{2} g_1 G_{1 \mu} +
  \frac{1 + \gamma_5}{2} g_0 G_{0 \mu} \right\} \chi^{\text{e}} . 
  \label{Delem}
\end{eqnarray}
Here again, it is assumed that the four-fermion interactions are suppressed by
another sufficiently large scale, and are henceforth neglected. At this stage,
there are no couplings between the composite and elementary sectors, as
evidenced in figure \ref{f:elem}, where the transformation properties of all
fields introduced up to now are summarized. The full $\mathcal{O} \left( p^2
\right)$ lagrangian is indeed the sum of terms involving composite fields on
one side and elementary ones on the other
\begin{eqnarray}
  \mathcal{L}_2 & = & \left( \mathcal{L}_{\tmop{SB}} +\mathcal{L}_{\tmop{comp}
  . \tmop{ferm}.} \right) + \left( \mathcal{L}_{\tmop{gauge}}
  +\mathcal{L}_{\tmop{elem}. \tmop{ferm}.} \right) .  \label{L2}
\end{eqnarray}
Its invariance group is
\begin{eqnarray}
  S_{\tmop{natural}} & = & \tmop{SU} \left( 2 \right)_L \times \tmop{SU}
  \left( 2 \right)_R \times \tmop{SU} \left( 2 \right)_{G_0} \times \tmop{SU}
  \left( 2 \right)_{G_1} \times \mathrm{U} \left( 1 \right)_{B - L} . 
  \label{Snat}
\end{eqnarray}

The couplings between the Goldstone modes and the elementary gauge fields are
introduced through constraints enforced via spurions, following the formalism
developed at length in {\cite{Hirn:2004ze}}. For convenience, we reproduce
here the main steps. The spurions link the $\tmop{SU} \left( 2 \right)_L
\times \tmop{SU} \left( 2 \right)_R \times \mathrm{U} \left( 1 \right)_{B -
L}$ transformations on one side with the $\tmop{SU} \left( 2 \right)_{G_0}
\times \tmop{SU} \left( 2 \right)_{G_1}$ gauge group on the other side. We
introduce the two-by-two matrices $X$ and $\tilde{Y}$ and the complex doublet
$\phi$. $X$ is taken to satisfy the reality condition
\begin{eqnarray}
  X & = & \tau^2 X^{\ast} \tau^2 .  \label{reality}
\end{eqnarray}
The transformation properties of the spurions are
\begin{eqnarray}
  X & \longmapsto & RXG_1^{\dag},  \label{trsf-X}\\
  \tilde{Y} & \longmapsto & G_0  \tilde{Y} L^{\dag}, \\
  \phi & \longmapsto & G_0 \mathe^{\mathi \frac{\alpha^0}{2}} \phi . 
\end{eqnarray}
Conditions of covariant constancy
\begin{eqnarray}
  D_{\mu} X & = & \partial_{\mu} X - \mathi R_{\mu} X + \mathi g_1 XG_{1 \mu}
   =  0,  \label{c1}\\
  D_{\mu} \tilde{Y} & = & \partial_{\mu} \tilde{Y} - \mathi g_0 G_{0 \mu} 
  \tilde{Y} + \mathi \tilde{Y} L_{\mu}  =  0, \\
  D_{\mu} \phi & = & \partial_{\mu} \phi - \mathi g_0 G_{0 \mu} \phi + \mathi
  \frac{B_{\mu}^0}{2} \phi  =  0,  \label{c2}
\end{eqnarray}
are imposed on the spurions. These constraints imply the correct vacuum
alignment of gauge connections and the coupling of gauge fields with Goldstone
bosons. The solution to the constraints (\ref{c1}-\ref{c2}) was given in
{\cite{Hirn:2004ze}}: it is possible to find a gauge {\emdash}which we will
call {\tmem{standard gauge}} {\emdash} where the spurions reduce to four real
constants and one phase~$\varphi$, yielding
\begin{eqnarray}
  \left. X \right|_{\text{s.g.}} & = & \xi \left(\begin{array}{cc}
    1 & 0\\
    0 & 1
  \end{array}\right),  \label{s1}\\
  \left. \tilde{Y} \right|_{\text{s.g.}} & = & \mathe^{\mathi \varphi} 
  \left(\begin{array}{cc}
    \eta_1 & 0\\
    0 & \eta_2
  \end{array}\right), \\
  \left. \phi \right|_{\text{s.g.}} & = & \left(\begin{array}{c}
    \zeta\\
    0
  \end{array}\right) .  \label{s2}
\end{eqnarray}
The constants $\xi$, $\eta_1$, $\eta_2$ and $\zeta$ will be considered as
small parameters. For notation purposes, we introduce the power counting
\begin{eqnarray}
  \xi, \eta_1, \eta_2 & = & \mathcal{O} \left( \epsilon \right) . 
\end{eqnarray}
The spurion $\zeta$ is likely to be even smaller {\cite{Hirn:2004ze}}, and we
count it separately. In the sequel, the phase $\varphi$ does not play any
physical role: it can be eliminated by appropriate redefinitions.

The standard gauge in which (\ref{s1}-\ref{s2}) hold is reached by making use
of an $\tmop{SU} \left( 2 \right)_R \times \mathrm{U} \left( 1 \right)_{B -
L}$ transformation, together with a $\tmop{SU} \left( 2 \right)_{G_0} \times
\tmop{SU} \left( 2 \right)_L / \mathrm{U} \left( 1 \right)_{L + G_0, \tau^3}$
transformation, leaving us with an $\tmop{SU} \left( 2 \right)_{G_1} \times
\mathrm{U} \left( 1 \right)_{L + G_0, \tau^3}$ gauge freedom. Indeed, in this
standard gauge, the connections are identified as a consequence of the
constraints (\ref{c1}-\ref{c2}) as follows
\begin{eqnarray}
  \left. R_{\mu} \right|_{\text{s.g.}} & = & g_1 G_{1 \mu},  \label{id1}\\
  \left. L^{1, 2}_{\mu} \right|_{\text{s.g.}} & = & G_{0 \mu}^{1, 2}
   =  0, \\
  \left. L^3_{\mu} \right|_{\text{s.g.}} & = & \left. B^0_{\mu}
  \right|_{\text{s.g.}}  =  g_0 G^3_{0 \mu}
   =  g_0 b^0_{\mu} .  \label{id3}
\end{eqnarray}
In the last equation, a new notation $b_{\mu}^0$ for the $\mathrm{U} \left( 1
\right)$ gauge field has been introduced. In summary, the spurions $\tilde{Y},
X, \phi$ have enabled us to reduce the symmetry from $S_{\tmop{natural}}$ down
to
\begin{eqnarray}
  S_{\tmop{reduced}} & = & \tmop{SU} \left( 2 \right) \times \mathrm{U} \left(
  1 \right)_Y . 
\end{eqnarray}
$S_{\tmop{reduced}}$ is actually the residual the symmetry of the solution of
the constraints (\ref{c1}-\ref{c2}).

\DOUBLEFIGURE{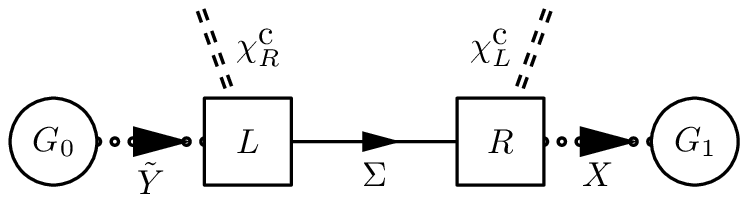, width=.45\textwidth}{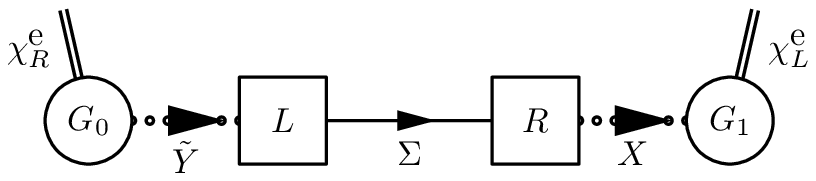, width=.45\textwidth}{Spurion
  couplings in the presence of composite fermions.\label{f:spu2}}{Spurion
  couplings in the presence of elementary fermions.\label{f:spu3}}

The transformation properties of all fields introduced up to now are
summarized in Figure \ref{f:spu2} and \ref{f:spu3}, where both cases with
composite and elementary fermions are presented separately for convenience.

We are now in a position to write down the lowest-order terms in the full
effective lagrangian. At leading order $\mathcal{O} \left( p^2 \epsilon^0
\zeta^0 \right)$, the lagrangian actually coincides with (\ref{L2}). The
interactions appear through the constraints (\ref{c1}-\ref{c2})
\begin{eqnarray}
  \mathcal{L}^{( 2, 0, 0 )} & = & \left. \left( \mathcal{L}_{\tmop{SB}}
  +\mathcal{L}_{\tmop{comp}. \tmop{ferm}.} +\mathcal{L}_{\tmop{gauge}}
  +\mathcal{L}_{\tmop{elem}. \tmop{ferm}.} \right)
  \right|_{\tmop{constraints}} .  \label{Op2-lag}
\end{eqnarray}
Terms invariant under $S_{\tmop{natural}}$ (\ref{Snat}) other than those
included in (\ref{Op2-lag}) will involve additional powers of momentum (or
coupling constants) or powers of the spurions, and are therefore suppressed.
The couplings are best evidenced by writing down the covariant derivative for
the Goldstone fields and then injecting the solution to the constraints, given
in (\ref{id1}-\ref{id3}). One finds
\begin{eqnarray}
  \nabla_{\mu} \Sigma & = & \left. D_{\mu} \Sigma \right|_{\text{s.g.}}
  = \partial_{\mu} \Sigma - \mathi g_0 b^0_{\mu}  \frac{\tau^3}{2} \Sigma
  + \mathi g_1 \Sigma G_{1 \mu} .  \label{c-der}
\end{eqnarray}
Performing the same manipulation for the covariant derivatives acting on both
composite~(\ref{Dcomp}) and elementary fermions (\ref{Delem}), the same
expression is obtained
\begin{eqnarray}
  \nabla_{\mu} \chi_L & = & \left. D_{\mu} \chi_L \right|_{\text{s.g.}}
   = \partial_{\mu} \chi_L - \mathi \left( g_1 G_{1 \mu} + g_0  \frac{B -
  L}{2} b^0_{\mu} \right) \chi_L, \\
  \nabla_{\mu} \chi_R & = & \left. D_{\mu} \chi_R \right|_{\text{s.g.}}
  = \partial_{\mu} \chi_R - \mathi g_0  \left( \frac{\tau^3}{2} + \frac{B
  - L}{2} \right) b^0_{\mu} \chi_R, 
\end{eqnarray}
where we have removed the superscripts $\text{e}$ and $\text{c}$. This is
indeed the desired result: we recognize SM-like covariant derivatives up to a
change in notation. This however does not imply that the couplings of the
composite fermions to the elementary gauge fields are identical to those of
the SM fermions, since the terms appearing with the free multiplicative factors
$\delta_L, \delta_R$ are allowed at leading order for composite fermions
(\ref{ar}).

Writing the $\mathcal{O} \left( p^2 \epsilon^0 \zeta^0\right)$ lagrangian
(\ref{Op2-lag}) in the same standard gauge where~(\ref{s1}-\ref{s2})
hold, we obtain
\begin{eqnarray}
  \left. \mathcal{L}^{( 2, 0, 0 )} \right|_{\text{s.g.}} & = & \frac{f^2}{4} 
  \left\langle \nabla_{\mu} \Sigma \nabla^{\mu} \Sigma^{\dag} \right\rangle -
  \frac{1}{2}  \left\langle G_{1 \mu \nu} G_1^{\mu \nu} \right\rangle -
  \frac{1}{4} b^0_{\mu \nu} b^{0 \mu \nu} + \mathi \overline{\chi^{\text{e}}}
  \gamma^{\mu} \nabla_{\mu} \chi^{\text{e}} \nonumber\\
  & + & \mathi \overline{\chi^{\text{c}}} \gamma^{\mu} \nabla_{\mu}
  \chi^{\text{c}} - m \left( \overline{\chi^{\text{c}}_L} \Sigma^{\dag}
  \chi^{\text{c}}_R + \overline{\chi^{\text{c}}_R} \Sigma \chi^{\text{c}}_L
  \right) \nonumber\\
  & + & \delta_L  \overline{\chi^{\text{c}}_L} \gamma^{\mu}  \left( \mathi
  \Sigma^{\dag} \partial_{\mu} \Sigma + g_0 b_{\mu}^0 \Sigma^{\dag} 
  \frac{\tau^3}{2} \Sigma - g_1 G_{1 \mu} \right) \chi^{\text{c}}_L
  \nonumber\\
  & + & \delta_R  \overline{\chi^{\text{c}}_R} \gamma^{\mu}  \left( \mathi
  \Sigma \partial_{\mu} \Sigma^{\dag} + g_1 \Sigma G_{1 \mu} \Sigma^{\dag} -
  g_0 b_{\mu}^0  \frac{\tau^3}{2}  \right) \chi^{\text{c}}_R,  \label{c-lag}
\end{eqnarray}
with $b^0_{\mu \nu}$ a standard $\mathrm{U} \left( 1 \right)$ field-strength
\begin{eqnarray}
  b^0_{\mu \nu} & = & \partial_{\mu} b^0_{\nu} - \partial_{\nu} b^0_{\mu} . 
\end{eqnarray}
The main specificity of composite fermions at this level comes from the
non-standard $\delta_L, \delta_R$ terms. It would seem that fermions with such
couplings are ruled out by experiments testing the universality of left-handed
couplings as well as couplings of right-handed fermions to the~$W^{\pm}$.
However, one may entertain the view that there may be such fermions in
addition to the three known generations. They must then be heavier in order to
have eluded detection at accelerators until now, and do not necessarily
constitute a full generation in the usual sense: the sum of $B - L$ over these
composite fermions may be non-vanishing {\footnote{We will come back to the
value of the trace of $B - L$ for elementary fermions in section \ref{sec:
elementary}.}}. As it happens, such composite fermions naturally come out
heavier than the elementary ones: this is evidenced by the fact that we can
build a mass term for them that does not involve spurions, as shown in
(\ref{m}), whereas mass terms for elementary fermions would be suppressed by
two powers of spurions {\cite{Hirn:2004ze}}. On the other hand,
mass-splittings appear in both cases only at $\mathcal{O} \left( p^1
\epsilon^2 \zeta^0 \right)$ since they require the spurion $\tilde{Y}$, meaning that
the ratio of splittings to the mean mass is smaller for composite fermions
doublets than for elementary ones. Of course, the two types of fermions may
also mix since multiplication of one type of field by the appropriate spurion
yields a field transforming as the other type of fermion. Such a mixing is a
higher-order effect in the spurion expansion, and is expected to be small. The
distinction between elementary and composite fermions therefore holds order-by-order in
the spurion expansion.

Except for the non-standard couplings $\delta_L$, $\delta_R$ of the composite
fermions (or in the absence of the latter), the lagrangian (\ref{c-lag}) leads
to EWSB with the same tree-level relations as in the SM without physical Higgs
particle~{\cite{Hirn:2004ze}}.

\section{Anomalies} \label{s:anomalies}

In this section, we discuss the anomaly-matching between the LEET and the
underlying techni-theory. We perform this analysis in the general case, taking
into account composite fermions if present. Since we will be dealing with the
matching between the techni-theory and its low-energy representation, the
introduction of the gauge symmetry $\tmop{SU} \left( 2 \right)_{G_0} \times
\tmop{SU} \left( 2 \right)_{G_1}$ can be relegated to a later stage: the
corresponding gauge fields only play the role of spectators. The same is true
of the spurions. We can thus use a description involving only the $\tmop{SU}
\left( 2 \right)_L \times \tmop{SU} \left( 2 \right)_R \times \mathrm{U}
\left( 1 \right)_{B - L}$ symmetry: we  focus on the relation between
the generating functional $\Gamma \left[ L_{\mu}, R_{\mu}, B^0_{\mu} \right]$
defined in (\ref{aa}) and its expression~(\ref{rep}) in terms of the low-energy variables.

\subsection{Anomalies of the techni-theory for low-energy symmetries}
\label{sec: TT}

The Noether currents of the techni-theory $J_L^{a \mu}, J_R^{a \mu}, J_{B -
L}^{\mu}$ were introduced in (\ref{aa}): they  correspond to the global
symmetries of the techni-theory that operate on the low-energy fields. We
study their anomalous Green's functions, denoting by $t$ a generic
transformation in the group of low-energy symmetries of the techni-theory
$\tmop{SU} \left( 2 \right)_L \times \tmop{SU} \left( 2 \right)_R \times
\mathrm{U \left( 1 \right)}_{B - L}$. We parameterize the transformation $t$
by the seven functions $\alpha = \alpha^a \tau^a / 2$, $\beta = \beta^a \tau^a
/ 2$ and~$\alpha^0$ as
\begin{eqnarray}
  t & = & \left( L = \mathe^{\mathi \left( \alpha - \beta \right)}, R =
  \mathe^{\mathi \left( \alpha + \beta \right)}, \alpha^0 \right) . 
  \label{ab}
\end{eqnarray}
Under a transformation $t$, the sources are modified according to
\begin{eqnarray}
  L_{\mu} & \longmapsto & \text{}^t L_{\mu} = LL_{\mu} L^{\dag} + \mathi L
  \partial_{\mu} L^{\dag},  \label{ae}\\
  R_{\mu} & \longmapsto & \text{}^t R_{\mu} = RR_{\mu} R^{\dag} + \mathi R
  \partial_{\mu} R^{\dag},  \label{af}\\
  B^0_{\mu} & \longmapsto & \text{}^t B^0_{\mu} = B^0_{\mu} - \partial_{\mu}
  \alpha^0 .  \label{ag}
\end{eqnarray}
At this point, it is useful to introduce differential operators acting on the
sources in order to reproduce the above transformations. However, in addition
to these sources, we have introduced the set of Goldstone fields $\left\{
\pi^a | a = 1, 2, 3 \right\}$ in order to reproduce the low-energy
singularities of the theory. These fields are collected in the unitary matrix
$\Sigma$
\begin{eqnarray}
  \Sigma & = & \mathe^{\mathi \frac{\pi^a \tau^a}{f}}  =
   \cos \left( \hat{\pi} \right) + \mathi \frac{\pi}{f \hat{\pi}}
  \sin \left( \hat{\pi} \right),  \label{pi}
\end{eqnarray}
where we have defined $\pi = \pi^a \tau^a$ and $\hat{\pi} = \sqrt{\pi^a \pi^a}
/ f$. We are therefore interested in a definition of the differential operator
that also acts directly on these Goldstone fields~{\cite{Wess:1971yu,Weinberg:1996kr}} {\emdash}as opposed to it operating via the equations of
motion {\cite{Gasser:1985gg}}. In order to reproduce the transformation of the
Goldstone modes (\ref{00000}) under $t$, we introduce the following
infinitesimal construction, explicited as the action of $D \left( \alpha,
\beta, \alpha^0 \right)$ on a generic functional $f \left[ \Sigma, L_{\mu},
R_{\mu}, B^0_{\mu} \right]$
\begin{eqnarray}
  &  & D \left( \alpha, \beta, \alpha^0 \right) f \left[ \Sigma, L_{\mu},
  R_{\mu}, B^0_{\mu} \right] \nonumber\\
  & = & \int \mathd x  \left\{ \alpha^0 \partial_{\mu} \frac{\delta f}{\delta
  B^0_{\mu} \left( x \right)}+ \left\{ - \left( \alpha^a + \beta^a \right)
  \partial_{\mu} + \mathi \left[ \alpha + \beta, R_{\mu} \right]^a \right\} 
  \frac{\delta f}{\delta R_{\mu}^a \left( x \right)} \right. \nonumber\\
&+& \left\{ - \left(
  \alpha^a - \beta^a \right) \partial_{\mu} + \mathi \left[ \alpha - \beta,
  L_{\mu} \right]^a \right\}  \frac{\delta f}{\delta L_{\mu}^a \left( x
  \right)}  \nonumber\\
  & + & \left. \left\{ \mathi \left[ \alpha, \pi \right]^a + \frac{f
  \hat{\pi}}{\sin \left( \hat{\pi} \right)}  \left( - \beta^a \cos \left(
  \hat{\pi} \right) + \beta^b  \frac{\pi^b \pi^a}{f^2  \hat{\pi}^2}  \left(
  \cos \left( \hat{\pi} \right) - \frac{1}{\hat{\pi}} \sin \left( \hat{\pi}
  \right) \right) \right) \right\}  \frac{\delta f}{\delta \pi^a \left( x
  \right)} \right\},  \label{ai}
\end{eqnarray}
One can check explicitly that (\ref{ai}) indeed yields the desired result
\begin{eqnarray}
  \mathe^{D \left( \alpha, \beta, \alpha^0 \right)} f \left[ \Sigma, L_{\mu},
  R_{\mu}, B^0_{\mu} \right] & = & f \left[ \text{}^t \Sigma, \text{}^t
  L_{\mu}, \text{}^t R_{\mu}, \text{}^t B^0_{\mu} \right] .  \label{ao}
\end{eqnarray}

With these definitions, we can perform the  Wess-Zumino construction
in our setting: since we will build upon the result in the sequel, we describe
this standard procedure. We use the differential operator $D ( \alpha, \beta,
\alpha^0 )$ of (\ref{ai}) to express the variation of the generating
functional $\Gamma \left[ L_{\mu}, R_{\mu}, B^0_{\mu} \right]$ defined in
(\ref{aa}) under a transformation applied on the techni-fields $T$. This
variation is determined by the content of the techni-theory. Assuming the
group of transformation $\tmop{SU} \left( 2 \right)_L \times \tmop{SU} \left(
2 \right)_R \times \mathrm{U \left( 1 \right)}_{B - L}$ to act linearly on the
techni-fermions, which constitute a subset of the fundamental variables $T$,
the action of the differential operator defined in (\ref{ai}) will yield a
standard Bardeen variation {\cite{Adler:1969gk,Bardeen:1969md,Adler:1969er,Fujikawa:1979ay,Fujikawa:1980eg,Balachandran:1982cs,Gasser:1984yg,Tsutsui:1989qr}}. This variation is therefore known up to a multiplicative
coefficient $k_{\tmop{TT}}$, which constitutes the only unknown here
\begin{eqnarray}
  D ( \alpha, \beta, \alpha^0 ) \Gamma \left[ L_{\mu}, R_{\mu}, B^0_{\mu}
  \right] & = & - k_{\tmop{TT}}  \int \mathd x \left\langle \beta \Omega
  \left[ L_{\mu}, R_{\mu}, B^0_{\mu} \right] \right\rangle,  \label{ap}
\end{eqnarray}
where a standard Bardeen functional appears
\begin{eqnarray}
  \Omega \left[ L_{\mu}, R_{\mu}, B^0_{\mu} \right] & = & \mathcal{O} \left(
  p^4 \epsilon^0 \zeta^0 \right) . 
\end{eqnarray}
For the group $\tmop{SU} \left( 2 \right)_L \times \tmop{SU} \left( 2
\right)_R \times \mathrm{U \left( 1 \right)}_{B - L}$, this functional assumes
the form {\cite{Kaiser:2000ck}}
\begin{eqnarray}
  \Omega \left[ L_{\mu}, R_{\mu}, B^0_{\mu} \right] & = & \frac{1}{16
  \mathpi^2} \varepsilon^{\mu \nu \rho \sigma} B^0_{\mu \nu}  \left(
  \partial_{\rho} L_{\sigma} + \partial_{\rho} R_{\sigma} - \mathi \left[
  R_{\rho}, L_{\sigma} \right] \right) .  \label{aq}
\end{eqnarray}
For a theory such as QCD, the value of the analogue of $k_{\tmop{TT}}$ is known: it is equal to $N_c / 3$ where~$N_c$ is the number of colors and the division by three is due to the baryon number of the quarks being equal to $1/3$ with the standard normalization of the $\tmop{U}\left(1\right)_{B - L}$ charges.
We will come back to the value of the constant $k_{\tmop{TT}}$ later in section \ref{sec: obstruction}.

One may check explicitly, using relation (\ref{ao}), that the Bardeen
functional in (\ref{aq}) verifies the Wess-Zumino integrability conditions
{\cite{Wess:1971yu}} expressing the commutation relation of the algebra
\begin{eqnarray}
  D \left( \frac{\tau^a}{2}, 0, 0 \right) \Omega^b \left[ L_{\mu}, R_{\mu},
  B^0_{\mu} \right] & = & \varepsilon^{abc} \Omega^c \left[ L_{\mu}, R_{\mu},
  B^0_{\mu} \right],  \label{aq1}\\
\varepsilon^{abc}  D \left( 0, \frac{\tau^a}{2}, 0 \right) \Omega^b \left[ L_{\mu}, R_{\mu},
  B^0_{\mu} \right] 
 & = & 0,  \label{aq2} \\
  D \left( 0, 0, \alpha^0 \right) \Omega \left[ L_{\mu}, R_{\mu}, B^0_{\mu}
  \right] & = & 0 .  \label{aq21}
\end{eqnarray}
These consistency conditions imply that the variation (\ref{ap}) can be
reproduced in the low-energy theory by an $\mathcal{O} \left( p^4 \epsilon^0 \zeta^0
\right)$ Wess-Zumino action $S_{\tmop{WZ}} \left[ \Sigma, L_{\mu}, R_{\mu},
B^0_{\mu} \right]$ depending on the Goldstone boson fields
\begin{eqnarray}
  D \left( \alpha, \beta, \alpha^0 \right) S_{\tmop{WZ}} \left[ \Sigma,
  L_{\mu}, R_{\mu}, B^0_{\mu} \right] & = & - \int \mathd x \left\langle \beta
  \Omega \left[ L_{\mu}, R_{\mu}, B^0_{\mu} \right] \right\rangle . 
  \label{aq4}
\end{eqnarray}
The appropriate action $S_{\tmop{WZ}} \left[ \Sigma, L_{\mu}, R_{\mu},
B^0_{\mu} \right]$ is found by integration of the differential equation~(\ref{aq4}). Under the boundary condition
\begin{eqnarray}
  S_{\tmop{WZ}} \left[ \mathbbm{1}, L_{\mu}, R_{\mu}, B^0_{\mu} \right] & = &
  0,  \label{aq5}
\end{eqnarray}
one performs the standard Wess-Zumino construction {\cite{Wess:1971yu}} to
obtain {\footnote{In this $\tmop{SU} \left( 2 \right)$ case, there is no need
to introduce a fifth compact dimension: the Wess-Zumino action is simply the
space-time integral of an ordinary lagrangian.}}
\begin{eqnarray}
  &  & S_{\tmop{WZ}} \left[ \Sigma, L_{\mu}, R_{\mu}, B^0_{\mu} \right]
  \nonumber\\
  & = & - \frac{1}{32 \mathpi^2}  \int \mathd x \varepsilon^{\mu \nu \rho
  \sigma} B^0_{\mu \nu}  \left\langle \Sigma R_{\rho} \Sigma^{\dag} L_{\sigma}
  - R_{\rho} L_{\sigma} - \mathi \Sigma^{\dag}  \left( \partial_{\rho} \Sigma
  \right) \left( R_{\sigma} + \Sigma^{\dag} L_{\sigma} \Sigma \right)
  \right\rangle \nonumber\\
  & - & \frac{1}{48 \mathpi^2}  \int \mathd x \varepsilon^{\mu \nu \rho
  \sigma} B^0_{\mu}  \left\langle \Sigma^{\dag} \left( \partial_{\nu} \Sigma
  \right) \Sigma^{\dag} \left( \partial_{\rho} \Sigma \right) \Sigma^{\dag}
  \left( \partial_{\sigma} \Sigma \right) \right\rangle .  \label{aq3}
\end{eqnarray}
From this formula, it can be checked explicitly, using the properties of the
differential operator (\ref{ai}) that we obtain the desired variation
(\ref{aq4}). Since this variation does not involve the fields $\pi^a$, and
since the integration measure over the Goldstone bosons is invariant under the
$\tmop{SU} \left( 2 \right)_L \times \tmop{SU} \left( 2 \right)_R \times
\mathrm{U \left( 1 \right)}_{B - L}$ symmetry, we immediately deduce the
variation of the generating functional. We thus get the usual result that this
standard Wess-Zumino action, when multiplied by the factor $k_{\tmop{TT}}$ and
added to a gauge-invariant action, reproduces the anomalous variation of the
generating functional (\ref{ap}).

This is the standard anomaly-matching
procedure. However, we also want to consider the case with fermionic
bound-states in the low-energy spectrum: these will in general imply an
additional variation of the action under the transformation $t$. The variation
will involve the Goldstone fields themselves and will depend on the
$\mathrm{U} \left( 1 \right)_{B - L}$ charges of the composite fermions as
well as on their couplings.

\subsection{Variation of the determinant for composite fermions} \label{sec:
composite}

On the LEET side, there is a contribution from composite fermions
{\cite{'tHooft:1979bh,Banks:1980sc,Frishman:1981dq}} to the anomalous variation of the
generating functional~$\Gamma \left[ L_{\mu}, R_{\mu}, B^0_{\mu} \right]$
under a transformation $t$. We consider a doublet of composite fermions
$\chi^{\text{c}}$ as introduced in section \ref{s:intr-comp}. To make things
more compact, we define the operator $\hat{\mathcal{D}}_{\mathrm{c}}$,
depending only on seven vector sources, by its action on $\chi^{\text{c}}$ as
follows
\begin{eqnarray}
  \hat{\mathcal{D}}_{\mathrm{c}} \left[ \hat{L}_{\mu}, \hat{R}_{\mu},
  B^0_{\mu} \right] \chi^{\text{c}} & = & \gamma^{\mu}  \left\{ \mathi
  \partial_{\mu} + \frac{B - L}{2} B^0_{\mu} + \frac{1 - \gamma_5}{2} 
  \hat{R}_{\mu} + \frac{1 + \gamma_5}{2}  \hat{L}_{\mu} \right\}
  \chi^{\text{c}} .  \label{at}
\end{eqnarray}
This allows us to write the lagrangian for composite fermions as
\begin{eqnarray}
  \mathcal{L}_{\tmop{comp}. \tmop{ferm}.} & = & \overline{\chi^{\text{c}}} 
  \hat{\mathcal{D}}_{\text{c}} \chi^{\text{c}}, 
\end{eqnarray}
provided we define the objects with carets by the relations
\begin{eqnarray}
  \hat{L}_{\mu} & = & L_{\mu} + \mathi \delta_R \Sigma D_{\mu} \Sigma^{\dag}, 
  \label{av}\\
  \hat{R}_{\mu} & = & R_{\mu} + \mathi \delta_L \Sigma^{\dag} D_{\mu} \Sigma, 
  \label{aw}
\end{eqnarray}
and we treat the Yukawa term as a perturbation. Given that the transformation
properties for the quantities with carets are just the same as for the
original ones
\begin{eqnarray}
  \hat{L}_{\mu} & \longmapsto & \text{}^t \hat{L}_{\mu} = L \hat{L}_{\mu}
  L^{\dag} + \mathi L \partial_{\mu} L^{\dag},  \label{az}\\
  \hat{R}_{\mu} & \longmapsto & \text{}^t \hat{R}_{\mu} = R \hat{R} _{\mu}
  R^{\dag} + \mathi R \partial_{\mu} R^{\dag},  \label{ba}
\end{eqnarray}
one can derive the following expression for the $\mathcal{O} \left( p^4
\epsilon^0 \zeta^0 \right)$ variation of the fermion determinant under a generic
transformation $t$
\begin{eqnarray}
  - \mathi D ( \alpha, \beta, \alpha^0 ) \ln \tmop{Det}
  \hat{\mathcal{D}}_{\mathrm{c}} \left[ \hat{L}_{\mu}, \hat{R}_{\mu},
  B^0_{\mu} \right] & = & \left( B - L \right) \int \mathd x \left\langle
  \beta \Omega \left[ \hat{L}_{\mu}, \hat{R}_{\mu}, B^0_{\mu} \right]
  \right\rangle .  \label{bb}
\end{eqnarray}
In this result, the same Bardeen functional as in (\ref{ap}) occurs, but with
different arguments. It can be rewritten as
\begin{eqnarray}
  \Omega \left[ \hat{L}_{\mu}, \hat{R}_{\mu}, B^0_{\mu} \right] & = & \Omega
  \left[ L_{\mu}, R_{\mu}, B^0_{\mu} \right] + \Omega_{\delta_L, \delta_R}
  \left[ \Sigma, L_{\mu}, R_{\mu}, B^0_{\mu} \right],  \label{bb1}
\end{eqnarray}
where $\Omega_{\delta_L, \delta_R}$ vanishes when both $\delta_L$ and
$\delta_R$ are set equal to zero
\begin{eqnarray}
  \Omega_{\delta_L, \delta_R} \left[ \Sigma, L_{\mu}, R_{\mu}, B^0_{\mu}
  \right] & = & \frac{1}{16 \mathpi^2} \varepsilon^{\mu \nu \rho \sigma}
  B^0_{\mu \nu}  \left\{ \delta_R  \left( \mathi \partial_{\rho} \left( \Sigma
  D_{\sigma} \Sigma^{\dag} \right) + \left[ R_{\rho}, \Sigma D_{\sigma}
  \Sigma^{\dag} \right] \right) \right. \nonumber\\
  & + & \delta_L  \left( \mathi \partial_{\rho} \left( \Sigma^{\dag}
  D_{\sigma} \Sigma \right) + \left[ L_{\rho}, \Sigma^{\dag} D_{\sigma} \Sigma
  \right] \right) \nonumber\\
  & + & \left. \mathi \delta_L \delta_R  \left[ \Sigma^{\dag} D_{\rho}
  \Sigma, \Sigma D_{\sigma} \Sigma^{\dag} \right] \right\} .  \label{bc}
\end{eqnarray}
One can see, from the transformation properties for $\hat{L}_{\mu}$ and
$\hat{R}_{\mu}$ given in (\ref{az}) and (\ref{ba}), that the Bardeen variation
in (\ref{bb}) will also satisfy the Wess-Zumino integrability conditions as in
(\ref{aq1}-\ref{aq21}), although this time the conditions themselves involve
the direct action of the operator $D$ on the $\pi^a$ fields. In fact the
non-standard variation $\Omega_{\delta_L, \delta_R} \left[ \Sigma, L_{\mu},
R_{\mu}, B^0_{\mu} \right]$ given in (\ref{bc}) satisfies the consistency
conditions on its own as well. Note that this remains true whatever the value
of the parameters $\delta_L$ and $\delta_R$. This means that we will be
able to build a non-standard Wess-Zumino lagrangian reproducing the additional
non-standard variation $\Omega_{\delta_L, \delta_R} \left[ \Sigma, L_{\mu},
R_{\mu}, B^0_{\mu} \right]$ separately, which we will need to do later. One
can exhibit the explicit expression for such an $\mathcal{O} \left( p^4
\epsilon^0 \zeta^0 \right)$ lagrangian $S_{\delta_L, \delta_R} \left[ \Sigma, L_{\mu},
R_{\mu}, B^0_{\mu} \right]$ according to
\begin{eqnarray}
  S_{\delta_L, \delta_R} \left[ \Sigma, L_{\mu}, R_{\mu}, B^0_{\mu} \right] &
  = & - \frac{1}{32 \mathpi^2}  \int \mathd x \varepsilon^{\mu \nu \rho
  \sigma} B^0_{\mu \nu}  \left\langle \mathi \left( \delta_L L_{\rho} -
  \delta_R R_{\rho} \right)  \left( \Sigma^{\dag} D_{\sigma} \Sigma + \Sigma
  D_{\sigma} \Sigma^{\dag} \right)  \right. \nonumber\\
  & - & \left( \delta_L - \delta_R \right) \partial_{\rho} \Sigma^{\dag}
  D_{\sigma} \Sigma \nonumber\\
  & + & \left. \delta_L \delta_R  \left( \Sigma^{\dag} D_{\rho} \Sigma \Sigma
  D_{\sigma} \Sigma^{\dag} - D_{\rho} \Sigma^{\dag} D_{\sigma} \Sigma \right)
  \right\rangle,  \label{bq}
\end{eqnarray}
from which it can be verified explicitly that the appropriate variation is
recovered
\begin{eqnarray}
  D \left( \alpha, \beta, \alpha^0 \right) S_{\delta_L, \delta_R} \left[
  \Sigma, L_{\mu}, R_{\mu}, B^0_{\mu} \right] & = & - \int \mathd x
  \left\langle \beta \Omega_{\delta_L, \delta_R} \left[ \Sigma,
  L_{\mu}, R_{\mu}, B^0_{\mu} \right] \right\rangle .  \label{br}
\end{eqnarray}
Notice that we have chosen $S_{\delta_L, \delta_R}$ in (\ref{bq}) to satisfy
the boundary condition
\begin{eqnarray}
  S_{\delta_L, \delta_R} \left[ \mathbbm{1}, L_{\mu}, R_{\mu}, B^0_{\mu}
  \right] & = & 0 .  \label{br1}
\end{eqnarray}

\subsection{Anomaly-matching} \label{sec: matching}

In section \ref{sec: TT} we obtained the total anomalous variation of the
generating functional $\Gamma \left[ L_{\mu}, R_{\mu}, B^0_{\mu} \right]$, as
deduced from the assumed properties of the techni-theory. In this same
section, we gave the form of the standard Wess-Zumino lagrangian to be added
to the invariant effective lagrangian so that the bosonic part of the LEET
reproduced the full variation. Then in section \ref{sec: composite}, we
considered the case where the LEET contains composite fermions in addition to
Goldstone bosons. We then determined the contribution of the variation of the
fermion determinant to the anomalous variation of the lagrangian (since this
variation in general contains $\Sigma$, and hence an integration over $\Sigma$
still has to be performed to obtain the action). In that case, on the LEET
side, it is the combination of the bosonic variation (the variation of a
Wess-Zumino-like term to be determined) and of the variation of the fermion
determinant (\ref{bb}), which has to match the total variation (\ref{ap}) derived from the
techni-theory. Finding the appropriate lagrangian for the LEET is
indeed what the anomaly-matching is about. In our case, this means
determining whether there are restrictions on the constants $\delta_L,
\delta_R$ and $B - L$ for the matching to be possible, and finding the form of
the Wess-Zumino-like lagrangian in the bosonic sector: we will perform this
matching at $\mathcal{O} \left( p^4 \epsilon^0 \zeta^0 \right)$, where the question
first arises.

With this in mind, we now consider how the variation of $\Gamma \left[
L_{\mu}, R_{\mu}, B^0_{\mu} \right]$ is reproduced in the effective theory.
Integrating over the composite fermions in (\ref{rep}), the low-energy
representation for the generating functional $\Gamma \left[ L_{\mu}, R_{\mu},
B^0_{\mu} \right]$ takes the form
\begin{eqnarray}
  \mathe^{\mathi \Gamma \left[ L_{\mu}, R_{\mu}, B^0_{\mu} \right]} & = & \int
  \mathd \left[ \Sigma \right] \mathe^{\mathi S_{\tmop{SB}} \left[ \Sigma,
  L_{\mu}, R_{\mu}, B^0_{\mu} \right] + \ln \tmop{Det}
  \hat{\mathcal{D}}_{\mathrm{c}} \left[ \hat{L}_{\mu}, \hat{R}_{\mu},
  B^0_{\mu} \right] +\mathcal{O} \left( p^4 \right)},  \label{ax}
\end{eqnarray}
with the obvious definition $S_{\tmop{SB}} = \int \mathd
x\mathcal{L}_{\tmop{SB}} $. The variation of $\Gamma \left[ L_{\mu},
R_{\mu}, B^0_{\mu} \right]$ given by expression (\ref{ax}) has to equal the
one given in (\ref{ap}). This is the anomaly-matching condition of 't Hooft
{\cite{'tHooft:1979bh}}, rephrased in the language of effective theories.

Focusing on the variation of the generating functional as reproduced by the
LEET~(\ref{ax}), we use successive rewritings to obtain
\begin{eqnarray}
&&  \mathe^{D ( \alpha, \beta, \alpha^0 )} \mathe^{\mathi \Gamma \left[ L_{\mu},
  R_{\mu}, B^0_{\mu} \right]} \nonumber\\
& = & \int \mathd \left[ \Sigma \right]
  \mathe^{\mathi S_{\tmop{SB}} \left[ \Sigma, \text{}^t L_{\mu}, \text{}^t
  R_{\mu}, \text{}^t B^0_{\mu} \right] + \ln \tmop{Det}
  \hat{\mathcal{D}}_{\mathrm{c}} \left[ \text{}^t \hat{L}_{\mu}, \text{}^t
  \hat{R}_{\mu}, \text{}^t B^0_{\mu} \right] +\mathcal{O} \left( p^4 \right)}
  \nonumber\\
  & = & \int \mathd \left[ \Sigma \right] \mathe^{D ( \alpha, \beta, \alpha^0
  )} \mathe^{\mathi S_{\tmop{SB}} \left[ \Sigma, L_{\mu}, R_{\mu}, B^0_{\mu}
  \right] + \ln \tmop{Det} \hat{\mathcal{D}}_{\mathrm{c}} \left[
  \hat{L}_{\mu}, \hat{R}_{\mu}, B^0_{\mu} \right] +\mathcal{O} \left( p^4
  \right)},  \label{bd}
\end{eqnarray}
where we have renamed the integration variables and then used invariance of
the Goldstone boson measure. This is in fact what allowed us to move the
differential operator $D ( \alpha, \beta, \alpha^0 )$ inside the functional
integral, even though it also acts on the pion fields. Some definitions are in
order: the action for the Goldstone bosons $S_{\tmop{SB}} \left[ \Sigma,
L_{\mu}, R_{\mu}, B^0_{\mu} \right]$ is split into two parts
\begin{eqnarray}
  \text{$S_{\tmop{SB}} \left[ \Sigma, L_{\mu}, R_{\mu}, B^0_{\mu} \right]$} &
  = & S_{\tmop{inv}} \left[ \Sigma, L_{\mu}, R_{\mu}, B^0_{\mu} \right] +
  \hat{S}_{\tmop{WZ}} \left[ \Sigma, L_{\mu}, R_{\mu}, B^0_{\mu} \right], 
  \label{be}
\end{eqnarray}
where $S_{\tmop{inv}}$ is invariant under a generic $t$ transformation
\begin{eqnarray}
  D ( \alpha, \beta, \alpha^0 ) S_{\tmop{inv}} \left[ \Sigma, L_{\mu},
  R_{\mu}, B^0_{\mu} \right] & = & 0,  \label{bf}
\end{eqnarray}
while $\hat{S}_{\tmop{WZ}}$ is not. We still have to determine what its
variation should be, depending on $\delta_L, \delta_R$ and $B - L$, in order
to satisfy the requirement of anomaly-matching, if this requirement can be
fulfilled at all. The ambiguity in the splitting (\ref{be}) is removed
provided we adopt a boundary condition such as
\begin{eqnarray}
  \hat{S}_{\tmop{WZ}} \left[ \mathbbm{1}, L_{\mu}, R_{\mu}, B^0_{\mu} \right]
  & = & 0,  \label{bg}
\end{eqnarray}
to pin down the expression for $\hat{S}_{\tmop{WZ}}$, which we are looking
for: the anomaly-matching will be satisfied if we can solve for
$\hat{S}_{\tmop{WZ}}$. In connection with this problem, we point out that a
related analysis has been performed in geometrical language in
{\cite{Alvarez-Gaume:1985yb}}.

Retaining terms in (\ref{bd}) that are linear in the parameters of the
transformation~$\alpha, \beta, \alpha^0$ and using the low-energy
representation for $\Gamma \left[ L_{\mu}, R_{\mu}, B^0_{\mu} \right]$, we
arrive at
\begin{eqnarray}
  0 & = & \int \mathd \left[ \Sigma \right] \mathe^{\mathi S_{\tmop{SB}}
  \left[ \Sigma, L_{\mu}, R_{\mu}, B^0_{\mu} \right] + \ln \tmop{Det}
  \hat{\mathcal{D}}_{\mathrm{c}} \left[ \hat{L}_{\mu}, \hat{R}_{\mu},
  B^0_{\mu} \right] +\mathcal{O} \left( p^4 \right)} \nonumber\\
&\times& \left\{ X \left[ \Sigma,
  L_{\mu}, R_{\mu}, B^0_{\mu} \right] +\mathcal{O} \left( p^6 \right)
  \right\},  \label{bj}
\end{eqnarray}
where $X$ is defined through
\begin{eqnarray}
  &  & X \left[ \Sigma, L_{\mu}, R_{\mu}, B^0_{\mu} \right] \nonumber\\
  & = & D ( \alpha, \beta, \alpha^0 ) \left\{ \hat{S}_{\tmop{WZ}} \left[
  \Sigma, L_{\mu}, R_{\mu}, B^0_{\mu} \right] - \mathi \ln \tmop{Det}
  \hat{\mathcal{D}}_{\mathrm{c}} \left[ \hat{L}_{\mu}, \hat{R}_{\mu},
  B^0_{\mu} \right] \right\} \nonumber\\
  & - & D ( \alpha, \beta, \alpha^0 ) \Gamma \left[ L_{\mu}, R_{\mu},
  B^0_{\mu} \right] .  \label{bk}
\end{eqnarray}
Expanding (\ref{bj}) in powers of sources, we deduce by iteration that $X$
vanishes when considered as a power series in the sources. Therefore, for our
purposes
\begin{eqnarray}
  X \left[ \Sigma, L_{\mu}, R_{\mu}, B^0_{\mu} \right] & = & 0,  \label{bm1}
\end{eqnarray}
and, from the definition for $X \left[ \Sigma, L_{\mu}, R_{\mu}, B^0_{\mu}
\right]$ (\ref{bk}), we deduce what the variation of~$\hat{S}_{\tmop{WZ}}$
must be in order to satisfy the anomaly-matching condition. Indeed, injecting
the expression for the total variation from the techni-theory (\ref{ap}) and
the variation of the fermion determinant (\ref{bb}) into this definition for
$X \left[ \Sigma, L_{\mu}, R_{\mu}, B^0_{\mu} \right]$, we can express the
variation of $\hat{S}_{\tmop{WZ}} \left[ \Sigma, L_{\mu}, R_{\mu}, B^0_{\mu}
\right]$ under a generic transformation $t$ as follows
\begin{eqnarray}
  D ( \alpha, \beta, \alpha^0 ) \hat{S}_{\tmop{WZ}} \left[ \Sigma, L_{\mu},
  R_{\mu}, B^0_{\mu} \right] & = & - k_{\tmop{TT}}  \int \mathd x \left\langle
  \beta \Omega \left[ L_{\mu}, R_{\mu}, B^0_{\mu} \right] \right\rangle
  \nonumber\\
  & - & \left( B - L \right) \int \mathd x \left\langle \beta \Omega \left[
  \hat{L}_{\mu}, \hat{R}_{\mu}, B^0_{\mu} \right] \right\rangle .  \label{bn}
\end{eqnarray}
One may rewrite this using the separation (\ref{bb1}) between standard
variation $\Omega \left[ L_{\mu}, R_{\mu}, B^0_{\mu} \right]$ and non-standard
variation $\Omega_{\delta_L, \delta_R} \left[ L_{\mu}, R_{\mu}, B^0_{\mu}
\right]$. We have already discussed the Wess-Zumino integrability conditions
for both pieces in (\ref{bn}) in sections \ref{sec: TT} and \ref{sec:
composite}. The outcome was that it is possible to build a lagrangian
reproducing the variation represented by each of these two terms separately,
whatever the values of $\delta_L, \delta_R$. Here we only have to build a
linear combination of the corresponding Wess-Zumino lagrangians: the standard
one and the non-standard one. Indeed, a particular solution to this
non-standard Wess-Zumino problem is given by
\begin{eqnarray}
  &  & \hat{S}_{\tmop{WZ}} \left[ \Sigma, L_{\mu}, R_{\mu}, B^0_{\mu} \right]
  \nonumber\\
  & = & \left( k_{\tmop{TT}} + B - L \right) S_{\tmop{WZ}} \left[ \Sigma,
  L_{\mu}, R_{\mu}, B^0_{\mu} \right] + \left( B - L \right) S_{\delta_L,
  \delta_R} \left[ \Sigma, L_{\mu}, R_{\mu}, B^0_{\mu} \right] .  \label{bo}
\end{eqnarray}
The boundary conditions for both terms in the last equation are given in
(\ref{aq5}) and (\ref{br1}): they imply that our construction automatically
verifies the initial condition (\ref{bg}) we have chosen.

Thus, equation (\ref{bo}) indeed gives us an action $\hat{S}_{\tmop{WZ}}$
satisfying (\ref{bn}). This shows that the anomaly-matching condition can be
solved whatever the value of $\delta_L, \delta_R$ and $B - L$, and that there
are therefore no restrictions on these constants. In addition, we have
obtained the expression of the non-standard Wess-Zumino lagrangian to be added
to the bosonic part of the effective lagrangian in order for the total
variation of the generating functional to be recovered. Our result obviously
accounts for the case without composite fermions as well, as can be found by
setting $B - L = 0$ in equation (\ref{bo}). The low-energy representation of
the generating functional $\Gamma$ reads, with the various pieces introduced
above
\begin{eqnarray}
  &&\mathe^{\mathi \Gamma \left[ L_{\mu}, R_{\mu}, B^0_{\mu} \right]} \nonumber\\
& = & \int
  \mathd \left[ \Sigma \right] \mathe^{\mathi S_{\tmop{inv}} \left[ \Sigma,
  L_{\mu}, R_{\mu}, B^0_{\mu} \right] + \mathi \hat{S}_{\tmop{WZ}} \left[
  \Sigma, L_{\mu}, R_{\mu}, B^0_{\mu} \right] + \ln \tmop{Det}
  \mathcal{D}_{\mathrm{c}} \left[ \hat{L}_{\mu}, \hat{R}_{\mu}, B^0_{\mu}
  \right] +\mathcal{O} \left( p^4 \right)} .  \label{bs}
\end{eqnarray}

Coming back to the assumed transformation properties for the composite
fermions, we may also consider the case where the left and right-handed
fermions have transformation properties others than those given by
(\ref{00200}) and (\ref{00210}). Except for an interchange between the left and
right-handed fermions, however, other choices lead to a Dirac operator whose
determinant is invariant. Indeed, in such cases, the left and right-handed
fermions do not transform independently under a transformation $t$, and
therefore there can be no anomalous variation of the action under such a
transformation. This then entails that such fermion doublets do not
participate in the anomaly-matching.

\section{Anomaly-matching in a Higgs-less effective theory} \label{s:U-gauge}

If we were to study the anomaly-matching in a theory where the Goldstone modes
remained in the spectrum, the results of the previous section would be
sufficient: there we performed the anomaly-matching, even in the case where
there are composite fermions in the spectrum. However, this is not the end of
the story, when one is interested in the situation where the (dynamical) Higgs
mechanism occurs, resulting in the removal of all Goldstone modes from the
spectrum.

At this stage, the gauge fields $G_{0 \mu}, G_{1 \mu}$ introduced in section
\ref{s:spurions} and coupled to the other fields via spurions come into play.
Indeed, the spurions introduce the proper identification between the gauge
fields and the chiral sources $L_{\mu}$, $R_{\mu}$, $B_{\mu}^0$, which
appeared directly in the discussion of anomalies. The situation is analogous
to that of $\chi$PT, where the global anomalies derived from QCD have to be
reproduced by the LEET. This is the anomaly-matching discussed in the previous
section, which is  implemented by adding a Wess-Zumino term in the LEET, with
particular care paid to possible composite fermions. However, we are
interested in theories of EWSB where all Goldstone modes can be gauged away
and the electroweak gauge bosons get their masses. We will require that the
techni-theory be such that this indeed happens: the lagrangian can then be
rewritten in terms of variables not involving Goldstone fields anymore.
Consequently, we have to study field redefinitions related to gauge
transformations in order to determine whether the $\Sigma$ field can be
eliminated. Anomalies will play an important part in the discussion at
$\mathcal{O} \left( p^4 \right)$ and higher orders. As it turns out, the
conclusion will be that such a redefinition is impossible unless anomalies for
the low-energy symmetries of the techni-theory vanish, that is, $k_{\tmop{TT}}
= 0$. This is  a general condition on the techni-theory itself.

This result is the same whether or not the techni-theory produces composite
fermions. Indeed, neither $B - L$ nor the non-standard couplings $\delta_L,
\delta_R$ will ever appear in the manipulations to be performed:  using the result of the
anomaly-matching of section \ref{s:anomalies}, the question
is phrased in terms of the content of the techni-theory itself,
instead of the low-energy variables. Concerning elementary fermions,
the consequence will be that the trace of $B - L$ over these elementary
fermions has to vanish.

\subsection{Anomaly obstruction: the composite sector} \label{sec:
obstruction}

We focus on the definition of the unitary gauge using field transformations,
in order to check whether the $\pi^a$ fields can be eliminated from the
action. These redefinitions of course rely on the gauge invariance of the
action and we ask what is their effect in presence of anomalies, that is, if
the constant $k_{\tmop{TT}}$ appearing as a multiplicative factor in front of
the anomalous variation of the generating functional for the techni-theory
(\ref{ap}) is non-zero. We will again study this question at the first order
at which it appears, that is $\mathcal{O} \left( p^4 \epsilon^0 \zeta^0\right)$. Taking
into account possible composite fermions, but not elementary fermions in this
section, we can in fact once more work at the stage before the gauge fields 
$G_{0 \mu}, G_{1 \mu}$ are introduced, as we will see.

We work at the level where the fermions are integrated out to yield a
determinant, allowing us to focus at first on the bosons. We will briefly come
back to the field redefinitions for the composite fermions in section
\ref{s:triangles}. For the case at hand, we will consider the action~$S \left[
\Sigma, L_{\mu}, R_{\mu}, B^0_{\mu} \right]$
\begin{eqnarray}
  S \left[ \Sigma, L_{\mu}, R_{\mu}, B^0_{\mu} \right] & = & S_{\tmop{inv}}
  \left[ \Sigma, L_{\mu}, R_{\mu}, B^0_{\mu} \right] + \hat{S}_{\tmop{WZ}}
  \left[ \Sigma, L_{\mu}, R_{\mu}, B^0_{\mu} \right] \nonumber\\
  & - & \mathi \ln \tmop{Det} \hat{\mathcal{D}}_{\mathrm{c}} \left[
  \hat{L}_{\mu}, \hat{R}_{\mu}, B^0_{\mu} \right],  \label{bt}
\end{eqnarray}
which occurs in the low-energy expression for $\Gamma \left[ L_{\mu}, R_{\mu},
B^0_{\mu} \right]$ (\ref{bs}). Integration over the gauge fields will be
performed at a later stage. However, one has to keep in mind that the three
vector sources $R_{\mu}^a$ will be identified with the dynamical fields
$G^a_{1 \mu}$, hence a field redefinition involving these sources can absorb
the three Goldstone bosons, while the three sources $L_{\mu}^a$ only contain
one dynamical field $L_{\mu}^3 = g_0 G_{0 \mu}^3$, and therefore cannot be used for the
same purpose. The assumed disappearance of the Goldstone modes from the
spectrum means that the action can be rewritten in terms of only seven vector
sources, but from the reasoning above, we see that we need only redefine the
$R_{\mu}$ sources. We will explicitly use the following definition in order to
obtain a new set of vector variables invariant under the~$\tmop{SU} \left( 2
\right)_R$ symmetry
\begin{eqnarray}
  g_1 W_{\mu} & = & \mathi \Sigma D_{\mu} \Sigma^{\dag}  =
   \Sigma R_{\mu} \Sigma^{\dag} + \mathi \Sigma \partial_{\mu}
  \Sigma^{\dag} - L_{\mu},  \label{bw}
\end{eqnarray}
and perform the change of variables
\begin{eqnarray}
  \left\{ \Sigma, L_{\mu}, R_{\mu}, B^0_{\mu} \right\} & \longrightarrow & 
  \left\{ \Sigma, L_{\mu}, g_1 W_{\mu}, B^0_{\mu} \right\} .  \label{bu}
\end{eqnarray}
The possibility to define the action in terms of variables invariant under the
gauge symmetry is a consequence of a well-known aspect of the Higgs mechanism ---the screening of the gauge charges. The question is then whether, using
the general writing
\begin{eqnarray}
  S \left[ \Sigma, L_{\mu}, R_{\mu}, B^0_{\mu} \right] & = & f \left[ L_{\mu},
  g_1 W_{\mu}, B^0_{\mu} \right] + h \left[ \Sigma, L_{\mu}, g_1 W_{\mu},
  B^0_{\mu} \right],  \label{bv}
\end{eqnarray}
where the $h$ functional a priori depends on $\Sigma$, we find or not
\begin{eqnarray}
  h \left[ \Sigma, L_{\mu}, g_1 W_{\mu}, B^0_{\mu} \right] & = & 0 . 
\end{eqnarray}
Again, to avoid ambiguities in the separation (\ref{bv}), we impose the
condition
\begin{eqnarray}
  h \left[ \mathbbm{1}, L_{\mu}, g_1 W_{\mu}, B^0_{\mu} \right] & = & 0 . 
\end{eqnarray}
We first make use of the fact that (\ref{bw}) is formally a gauge
transformation. Indeed we can write
\begin{eqnarray}
  g_1 W_{\mu} & = & ^{\omega} R_{\mu} - L_{\mu},  \label{ca}
\end{eqnarray}
with the transformation parameter chosen to depend on the Goldstone fields in
the following manner
\begin{eqnarray}
  \omega & = & \left( L = \mathbbm{1}, R = \Sigma, \alpha^0 = 0 \right) . 
  \label{bz}
\end{eqnarray}
This would indeed be sufficient if only the invariant part in the action
(\ref{bt}) was present, as we have
\begin{eqnarray}
  S_{\tmop{inv}} \left[ \Sigma, L_{\mu}, R_{\mu}, B^0_{\mu} \right] & = &
  S_{\tmop{inv}} \left[ \mathbbm{1}, g_1 W_{\mu} + L_{\mu}, L_{\mu}, B^0_{\mu}
  \right] .  \label{by}
\end{eqnarray}
Note that we really are only performing field redefinitions and not fixing a
gauge {\cite{Gross:1972pv,Grosse-Knetter:1993nn}}, therefore there will be no
conflict with the fact that we have to choose a particular gauge to
later solve the constraints on the spurions, as done in section
\ref{s:spurions}: the gauge transformations are merely invoked to guide our
reasoning.

We can now study the consequences on the total action $S$. In fact, we first
consider the case of infinitesimal $\pi^a$ fields and use the result of the
anomaly-matching for an infinitesimal transformation
\begin{eqnarray}
  D \left( \alpha, \beta, \alpha^0 \right) S \left[ \Sigma, L_{\mu}, R_{\mu},
  B^0_{\mu} \right] & = & - k_{\tmop{TT}}  \int \mathd x \left\langle \beta
  \Omega \left[ L_{\mu}, R_{\mu}, B^0_{\mu} \right] \right\rangle . 
  \label{cc}
\end{eqnarray}
We next inject in this relation the infinitesimal version of the chiral
transformation (\ref{bz}) used above in order to eliminate the Goldstone
fields from the even intrinsic-parity terms. This transformation assumes the
form
\begin{eqnarray}
  \omega & = & \left( L = \mathbbm{1}, R = 1 + \mathi \frac{\pi}{f}
  +\mathcal{O} \left( \pi^2 \right), \alpha^0 = 0 \right),  \label{ce}
\end{eqnarray}
where the Goldstone fields $\pi$ were defined in (\ref{pi}). Injecting
(\ref{ce}) into (\ref{cc}) and reshuffling the terms, the result (\ref{cc})
can be recast in a form more useful for our purposes
\begin{eqnarray}
  S \left[ \Sigma, L_{\mu}, R_{\mu}, B^0_{\mu} \right] & = & S \left[
  \mathbbm{1}, L_{\mu}, g_1 W_{\mu} + L_{\mu}, B^0_{\mu} \right] \nonumber\\
  & + & \frac{k_{\tmop{TT}}}{2 f}  \int \mathd x \left\langle \pi \Omega
  \left[ L_{\mu}, g_1 W_{\mu} + L_{\mu}, B^0_{\mu} \right] \right\rangle
  +\mathcal{O} \left( \pi^2 \right) .  \label{cg}
\end{eqnarray}
We can identify the part independent on the Goldstone boson fields
\begin{eqnarray}
  f \left[ L_{\mu}, g_1 W_{\mu}, B^0_{\mu} \right] & = & S \left[ \mathbbm{1},
  L_{\mu}, g_1 W_{\mu} + L_{\mu}, B^0_{\mu} \right], 
\end{eqnarray}
whereupon we see that the power series of $h$ in powers of $\pi^a$ begins with
the linear term
\begin{eqnarray}
  h \left[ \Sigma, L_{\mu}, g_1 W_{\mu}, B^0_{\mu} \right] & = &
  \frac{k_{\tmop{TT}}}{2 f}  \int \mathd x \left\langle \pi \Omega \left[
  L_{\mu}, g_1 W_{\mu} + L_{\mu}, B^0_{\mu} \right] \right\rangle +\mathcal{O}
  \left( \pi^2 \right) .  \label{ch}
\end{eqnarray}
The only possibility for this functional to be independent of the Goldstone
modes is then to have
\begin{eqnarray}
  k_{\tmop{TT}} & = & 0,  \label{ci}
\end{eqnarray}
that is, if the symmetries of the techni-theory that are operative on the
low-energy degrees of freedom are anomaly-free. In that case only can we get
rid of the Goldstone fields via field redefinitions. We now show that this
remains true for finite values of the Goldstone fields.

If condition (\ref{ci}) is met by the techni-theory, we can show that the
field redefinition~(\ref{bw}) does eliminate the Goldstone boson field in the
finite case as well: we then have
\begin{eqnarray}
  S \left[ \mathbbm{1}, L_{\mu}, g_1 W_{\mu} + L_{\mu}, B^0_{\mu} \right] & =
  & \left. \mathe^{D \left( \alpha, \beta, 0 \right)} S \left[ \Sigma,
  L_{\mu}, R_{\mu}, B^0_{\mu} \right] \right|_{\left( \alpha, \beta \right) =
  \left( \frac{\pi}{2 f}, \frac{\pi}{2 f} \right)} \nonumber\\
  & = & S \left[ \Sigma, L_{\mu}, R_{\mu}, B^0_{\mu} \right],  \label{cj}
\end{eqnarray}
where, in the last step, we have used the absence of anomalous variation of $S
\left[ \Sigma, L_{\mu}, R_{\mu}, B^0_{\mu} \right]$ in (\ref{cc}) when
$k_{\tmop{TT}} = 0$. The functional $h \left[
\Sigma, L_{\mu}, g_1 W_{\mu}, B^0_{\mu} \right]$, defined in (\ref{bv}), is
then found to vanish, concluding our proof.

To summarize the main result of the present section, we recall that the
condition that the Goldstone modes do not appear in the spectrum requires the
absence of anomalies for the low-energy symmetries of the techni-theory:
$k_{\tmop{TT}}$ has to vanish. If there are light composite fermions,
the LEET must involve a Wess-Zumino-like term in order to compensate for the anomalous variation of the fermion determinant. We have shown that such a
Wess-Zumino term can be constructed whatever the value of $\delta_L, \delta_R$
and $B - L$. On the other hand, if there are no light composite fermions, then
there will be no Wess-Zumino term, since $k_{\tmop{TT}}$ has to vanish.

We have concluded that, for those Higgs-less models to which our effective description applies, the underlying theory has to be anomaly-free with respect to the global group $\tmop{SU}\left(2\right)_L \times \tmop{SU}\left(2\right)_R \times \tmop{U}\left(1\right)_{B-L}$, in order for the GBs to disappear from the spectrum. Considering applications to technicolor models, one must keep in mind that the larger $S_{\tmop{natural}}$ symmetry was not introduced in the literature. On the other hand, technicolor models are usually constructed to be anomaly-free under the electroweak gauge group $\tmop{SU}\left(2\right) \times \tmop{U}\left(1\right)_Y$~\cite{Farhi:1981xs}, which is the corresponding requirement.
In order for anomalies to vanish, one adds new particles, for instance strongly-interacting technileptons to balance the $B-L$ charges of the techniquarks. The global symmetry group is then larger than that of QCD with two massless flavors (i.e.  than $\tmop{SU}\left(2\right)_L \times \tmop{SU}\left(2\right)_R \times \tmop{U}\left(1\right)_{B-L}$).
One expects this modification to be reflected in the spectrum of (massive) bound states, and therefore the spectral functions should a priori behave differently. Therefore, the estimates for the value of the $S$ parameter (alternatively $L_{10}$), derived from QCD are not theoretically justified in the cases of interest for Higgs-less EWSB, since these parameters are directly linked with the vector and axial spectral functions~\cite{Peskin:1992sw}.

We stress again that, using the result from the anomaly-matching derived
previously, we did not have to deal with the couplings of possible composite
fermions: these simply did not show up in the reasoning. As far as the Higgs
mechanism is concerned, composite fermions may just as well be absent: the
derivation would be identical.

Actually, as far as anomalies are concerned, the main difference between
composite fermions (if they are present) and elementary ones are concerns the
trace of the quantum number $B - L$. Whereas for elementary fermions, $\sum
\left( B - L \right)$ has to vanish (see next section), for composite fermions
this is not necessary: composite fermions do not necessarily constitute a full
generation as the SM fermions do. There is in fact an additional distinction
stemming from the mass term (\ref{m}): compared to elementary fermions,
composite fermions are heavier (see {\cite{Hirn:2004ze}}) but have a
relatively smaller mass-splitting within the doublet. Obviously, these are
theoretical distinctions, and composite fermions might be mistaken for
elementary ones in experiments, especially if $\delta_L \simeq \delta_R \simeq
0$.

\subsection{Elementary fermions in the unitary gauge} \label{sec: elementary}

In this section, we consider elementary fermions, instead of the composite
fermions we have been dealing with up to now. For these fermions, there is no
anomaly-matching to be performed since the fundamental variables are assumed
to correspond directly to the physical degrees of freedom at low-energy, due
to the perturbative nature of their couplings. We thus focus on the question
of anomalies as an obstruction to the elimination of the Goldstone modes from
the action: as we shall see, the results are independent of our previous
discussion concerning composite fermions. Elementary fermions have been
studied in an exactly identical setting in {\cite{Hirn:2004ze}}, and we only
discuss additional aspects related to anomalies. We will find that the trace
of $B - L$ over the elementary fermions has to vanish for the action to be
independent of the Goldstone modes.

We define the Dirac operator $\mathcal{D}_{\text{e}}$
\begin{eqnarray}
  \mathcal{D}_{\text{e}} \chi^{\text{e}} & = & \mathi \gamma^{\mu} D_{\mu}
  \chi^{\text{e}}, 
\end{eqnarray}
such that the $\mathcal{O} \left( p^2 \right)$ lagrangian for elementary
fermions is given by
\begin{eqnarray}
  \mathcal{L}_{\tmop{elem}. \tmop{ferm}.} & = & \overline{\chi^{\text{e}}}
  \mathcal{D}_{\text{e}} \chi^{\text{e}} . 
\end{eqnarray}
The transformation properties of $\mathcal{D}_{\text{e}}$ under the symmetries
$\tmop{SU} \left( 2 \right)_L \times \tmop{SU} \left( 2 \right)_R$ of the
techni-theory are trivial given that the elementary fermions on which it is to
be applied are external to this theory, and we need only consider $\tmop{SU}
\left( 2 \right)_{G_0} \times \tmop{SU} \left( 2 \right)_{G_1} \times
\mathrm{U} \left( 1 \right)_{B - L}$ transformations. Under such a
transformation $\tau$ parameterized
\begin{eqnarray}
  \tau & = & \left( G_0 = \mathe^{\mathi \left( \theta - \varphi \right)}, G_1
  = \mathe^{\mathi \left( \theta + \varphi \right)}, \alpha^0 \right), 
  \label{tau}
\end{eqnarray}
the variation of the determinant is given by the following expression, where
we restrict to the first non-zero term, that is $\mathcal{O} \left( p^4
\epsilon^0 \right)$
\begin{eqnarray}
  &  & - \mathi D \left( \theta, \varphi, \alpha^0 \right) \ln \tmop{Det}
  \mathcal{D}_{\mathrm{e}} \left[ g_0 G_{0 \mu}, g_1 G_{1 \mu}, B^0_{\mu}
  \right] \nonumber\\
  & = & \left( B - L \right)  \int \mathd x \left\langle \varphi \Omega
  \left[ g_0 G_{0 \mu}, g_1 G_{1 \mu}, B^0_{\mu} \right] \right\rangle . 
  \label{ct}
\end{eqnarray}
The Bardeen functional $\Omega$ occurring here is the same as appeared
previously in (\ref{aq}) albeit with different functional arguments. We may
explicitly display its expression as
\begin{eqnarray}
  \Omega \left[ g_0 G_{0 \mu}, g_1 G_{1 \mu}, B^0_{\mu} \right] & = &
  \frac{1}{16 \mathpi^2} \varepsilon^{\mu \nu \rho \sigma} B^0_{\mu \nu} 
  \left( g_0 \partial_{\rho} G_{0 \sigma} + g_1 \partial_{\rho} G_{1 \sigma} -
  \mathi g_0 g_1  \left[ G_{0 \rho}, G_{1 \sigma} \right] \right) . 
  \label{cu}
\end{eqnarray}

We now turn to the field redefinitions required to obtain the unitary gauge
variables for the bosonic sector. Since the elementary fermions do not
transform under $\tmop{SU} \left( 2 \right)_R$ but rather under $\tmop{SU}
\left( 2 \right)_{G_1}$, we have to redefine the $G_{1 \mu}$ fields rather
than the $R_{\mu}$ sources as we did in (\ref{bw}) in order to absorb the
three Goldstone bosons. We would like to perform the field redefinitions at
the level of the full covariant lagrangian, that is, regardless of any gauge
choice. Notice that the real spurion $X$, connecting $R_{\mu}$ and $g_1 G_{1 \mu}$ can be
decomposed as
\begin{eqnarray}
  X & = & \xi ( x ) U ( x ), 
\end{eqnarray}
where $\xi \left( x \right)$ is a real function and $U \in \tmop{SU} \left( 2
\right)$. The appropriate redefinition of the fields~$G^a_{1 \mu}$ that
absorbs the matrix~$\Sigma$ into the three vector fields~$\mathcal{W}^a_{\mu}$
is then
\begin{eqnarray}
  g_1 \mathcal{W}_{\mu} & = & \mathi \Sigma UD_{\mu} \left( U^{\dag}
  \Sigma^{\dag} \right) \nonumber\\
  & = & \mathi \Sigma U \partial_{\mu} \left( U^{\dag} \Sigma^{\dag} \right)
  + g_1 \Sigma UG_{1 \mu} U^{\dag} \Sigma^{\dag} - L_{\mu},  \label{sec-redef}
\end{eqnarray}
instead of the definition previously given in (\ref{bw}). On the space of
solutions of the constraints $D_{\mu} X = 0$ the two definitions are actually
identical. Indeed, the constraint of covariant constancy applied on the
spurion $X$, results in $\xi$ being a constant and in
\begin{eqnarray}
  D_{\mu} U  \equiv  \partial_{\mu} U - \mathi R_{\mu}
  U + \mathi g_1 UG_{1 \mu} & = & 0, 
\end{eqnarray}
yielding
\begin{eqnarray}
  \mathcal{W}_{\mu} & = & W_{\mu} . 
\end{eqnarray}
This result is true as it stands, due to the constraints, and without any
gauge choice. We notice that the field redefinition (\ref{sec-redef}) is again
formally a gauge transformation, but this time it is a particular case of a
$\tau$ transformation as introduced in (\ref{tau}) {\emdash}as opposed to a~$t$ transformation (\ref{ab}){\emdash} with parameters depending on the
Goldstone and spurion fields
\begin{eqnarray}
  \lambda & = & \left( G_0 = \mathbbm{1}, G_1 = \Sigma U, \alpha^0 = 0
  \right),  \label{lam}
\end{eqnarray}
so that we can rewrite (\ref{sec-redef}) as
\begin{eqnarray}
  g_1 \mathcal{W}_{\mu} & = & g_1  \text{}^{\lambda} G_{1 \mu} - L_{\mu} . 
\end{eqnarray}

To study the consequences of the field redefinition (\ref{sec-redef}), we use
the outcome of the integration
\begin{eqnarray}
  \int \mathd \left[ \chi^{\text{e}} \right] \mathe^{\mathi \int \mathd x
  \overline{\chi^{\text{e}}} \mathcal{D}_{\mathrm{e}} \left[ g_0 G_{0 \mu},
  g_1 G_{1 \mu}, B^0_{\mu} \right] \chi^{\text{e}}} & = & \mathe^{\ln
  \tmop{Det} \mathcal{D}_{\mathrm{e}} \left[ g_0 G_{0 \mu}, g_1 G_{1 \mu},
  B^0_{\mu} \right]},  \label{cv}
\end{eqnarray}
and proceed as in section \ref{sec: obstruction}. We start with the case of an
infinitesimal $\pi$ field and infinitesimal $\kappa$ of the same order defined as
\begin{eqnarray}
  U & = & \mathe^{\mathi \frac{\kappa}{f}} . 
\end{eqnarray}
Application of an infinitesimal transformation then yields
\begin{eqnarray}
  &  & - \mathi D \left( \theta = \frac{\pi + \kappa}{2 f}, \varphi =
  \frac{\pi + \kappa}{2 f}, \alpha^0 = 0 \right) \ln \tmop{Det}
  \mathcal{D}_{\mathrm{e}} \left[ g_0 G_{0 \mu}, g_1 G_{1 \mu}, B^0_{\mu}
  \right] \nonumber\\
  & = & \frac{B - L}{2 f}  \int \mathd x \left\langle \left( \pi + \kappa
  \right) \Omega \left[ g_0 G_{0 \mu}, g_1 \mathcal{W}_{\mu} + L_{\mu},
  B^0_{\mu} \right] \right\rangle \nonumber\\
  & + & \mathcal{O} \left( \pi^2, \pi \kappa, \kappa^2 \right) .  \label{cw}
\end{eqnarray}
This shows that it is impossible to absorb the Goldstone boson fields in the
action unless the trace of $B - L$ over elementary fermions vanishes. To
deduce this, we have used the fact that there can be no cancellation between
(\ref{cw}) and (\ref{cg}): the anomalous variation of the determinant for
elementary fermions must vanish by itself in order to get rid of the Goldstone
modes. Thus we must impose for consistency
\begin{eqnarray}
  \sum_{\tmop{elementary} \tmop{fermions}} \left( B - L \right) & = & 0 . 
  \label{cw1}
\end{eqnarray}
One then checks that the field redefinitions do also work for finite
transformations. Usually, the condition (\ref{cw1}) derives from the
requirement of renormalizability of the theory~{\cite{Bouchiat:1972iq,Gross:1972pv,KorthalsAltes:1972aq,Georgi:1972bb}}, which is not of an
obvious relevance within the present framework of LEET. Nevertheless, this
relation reappears here as a necessary condition for the Higgs-less symmetry
breaking mechanism to occur.

Finally, in order to write the action in unitary gauge at the level of the
functional integral for fermions, we may, in addition to the field
redefinitions in the bosonic sector, introduce the following redefinition for
fermions
\begin{eqnarray}
  \psi^{\text{e}}_L & = & \text{}^{\lambda} \chi^{\text{e}}_L  =
   \Sigma U \chi^{\text{e}}_L,  \label{cy}\\
  \psi^{\text{e}}_R & = & \text{}^{\lambda} \chi^{\text{e}}_R  =
   \chi^{\text{e}}_R .  \label{cz}
\end{eqnarray}
This is in fact the definition implied by the transformation $\lambda$
(\ref{lam}). The new $\psi^{\text{e}}$ variables are invariant under the
$\tmop{SU} \left( 2 \right)_R \times \tmop{SU}\left(2\right)_{G_1}$ symmetry, as are the $W_{\mu}$ fields: this is
indeed the purpose of these redefinitions
\begin{eqnarray}
  \psi^{\text{e}}_L & \longmapsto & \text{}^t \psi^{\text{e}}_L  = \quad
  L \mathe^{- \mathi \frac{B - L}{2} \alpha^0} \psi^{\text{e}}_L, 
  \label{da}\\
  \psi^{\text{e}}_R & \longmapsto & \text{}^t \psi^{\text{e}}_R  = \quad
  G_0 \mathe^{- \mathi \frac{B - L}{2} \alpha^0} \psi^{\text{e}}_R .  \label{db}
\end{eqnarray}
Since the lagrangian for elementary fermions (\ref{cr}), as well as the
fermionic integration measure are both invariant under the transformation
$\lambda$, we have
\begin{eqnarray}
  \int \mathd \left[ \chi^{\text{e}} \right] \mathe^{\mathi
  \overline{\chi^{\text{e}}} \mathcal{D}_{\mathrm{e}} \left[ g_0 G_{0 \mu},
  g_1 G_{1 \mu}, B^0_{\mu} \right] \chi^{\text{e}}}  & = & \int \mathd \left[
  \psi^{\text{e}} \right] \mathe^{\mathi \overline{\psi^{\text{e}}}
  \mathcal{D}_{\mathrm{e}} \left[ g_0 G_{0 \mu}, g_1 \mathcal{W}_{\mu} +
  L_{\mu}, B^0_{\mu} \right] \psi^{\text{e}}} .  \label{dc1}
\end{eqnarray}
The conclusion is then that the lagrangian in the unitary gauge is directly
obtained via the replacement
\begin{eqnarray}
  \left\{ \chi^{\text{e}}, \Sigma, G_{1 \mu}, B^0_{\mu} \right\} &
  \longrightarrow & \left\{ \psi^{\text{e}}, \mathbbm{1},\mathcal{W}_{\mu} +
  \frac{1}{g_1} L_{\mu}, B^0_{\mu} \right\} .  \label{cx}
\end{eqnarray}
\section{Triangles in the unitary gauge} \label{s:triangles}

We now wish to introduce the corresponding redefinitions for composite
fermions in the unitary gauge. They should be such that the composite fermion
fields are invariant under~$\tmop{SU} \left( 2 \right)_R$. Therefore, we set,
using the $\omega$ transformation from (\ref{bz})
\begin{eqnarray}
  \psi^{\text{c}}_L & = & ^{\omega} \chi^{\text{c}}_L  =
   \Sigma \chi^{\text{c}}_L,  \label{cl}\\
  \psi^{\text{c}}_R & = & ^{\omega} \chi^{\text{c}}_R  =
   \chi^{\text{c}}_R,  \label{cm}
\end{eqnarray}
which indeed yields
\begin{eqnarray}
  \psi^{\text{c}}_L & \longmapsto & ^t \psi^{\text{c}}_L  =  L
  \mathe^{- \mathi \frac{B - L}{2} \alpha^0} \psi^{\text{c}}_L,  \label{cn}\\
  \psi^{\text{c}}_R & \longmapsto & ^t \psi^{\text{c}}_R  =  L
  \mathe^{- \mathi \frac{B - L}{2} \alpha^0} \psi^{\text{c}}_R,  \label{co}
\end{eqnarray}
as desired.

Of course, the determinant of the Dirac operator~$\mathcal{D}_{\text{c}}$ defined as $\mathcal{D}_{\text{c}}\left[\Sigma, L_{\mu}, R_{\mu}, B^0_{\mu} \right] = \hat{\mathcal{D}}_{\text{c}}\left[\hat{L}_{\mu}, \hat{R}_{\mu}, B^0_{\mu} \right]$  is not invariant under the~$\omega$ transformation.
However, due to the absence of
anomalies for the whole techni-theory ($k_{\tmop{TT}} = 0$), this variation is
compensated by that of the Wess-Zumino term. Hence we can rewrite (\ref{cj})
in the form
\begin{eqnarray}
  &  & \mathe^{\mathi S_{\tmop{inv}} \left[ \Sigma, L_{\mu}, R_{\mu},
  B_{\mu}^0 \right] + \mathi \hat{S}_{\tmop{WZ}} [ \Sigma, L_{\mu}, R_{\mu},
  B_{\mu}^0 ]}  \int \mathd \left[ \chi^{\text{c}} \right] \mathe^{\mathi \int
  \mathd x \overline{\chi^{\text{c}}} \mathcal{D}_{\text{c}} \left[ \Sigma,
  L_{\mu}, R_{\mu}, B^0_{\mu} \right] \chi^{\text{c}}} \nonumber\\
  & = & \mathe^{\mathi S_{\tmop{SB}} \left[ \mathbbm{1}, L_{\mu}, g_1 W_{\mu}
  + L_{\mu}, B^0_{\mu} \right]}  \int \mathd \left[ \psi^{\text{c}} \right]
  \mathe^{\mathi \int \mathd x \overline{\psi^{\text{c}}}
  \mathcal{D}_{\mathrm{c}} \left[ \mathbbm{1}, L_{\mu}, g_1 W_{\mu} + L_{\mu},
  B^0_{\mu} \right] \psi^{\text{c}} },  \label{ineq}
\end{eqnarray}
where we have used the boundary condition (\ref{bg}) for
$\hat{S}_{\tmop{WZ}}$. The outcome is again that we merely have to perform the
following replacement in the lagrangian
\begin{eqnarray}
  \left\{ \chi^{\text{c}}, \Sigma, L_{\mu}, R_{\mu},
  B^0_{\mu} \right\} & \longmapsto &  \left\{ \psi^{\text{c}}, \mathbbm{1}, L_{\mu}, g_1 W_{\mu} + L_{\mu}, B^0_{\mu}
  \right\}, 
\end{eqnarray}
in order to obtain the lagrangian in unitary gauge.

It is in fact quite natural to work in the unitary gauge for such effective
theories: this has the additional advantage that the Wess-Zumino-like term
vanishes in this gauge, due to the boundary condition (\ref{bg}). We therefore
use this gauge for the following discussion of anomalous triangular diagrams:
we can write
\begin{eqnarray}
&&  \int \mathd \left[ \Sigma \right] \mathe^{\mathi S \left[ \Sigma, L_{\mu},
  R_{\mu}, B^0_{\mu} \right]} \nonumber\\
& = & \int \mathd \left[ \psi^{\text{c}} \right]
  \mathe^{\mathi \int \mathd x\mathcal{L}_{\tmop{SB}} \left[ \mathbbm{1},
  L_{\mu}, g_1 W_{\mu} + L_{\mu}, B^0_{\mu} \right] +
  \overline{\psi^{\text{c}}} \mathcal{D}_{\mathrm{c}} \left[ \mathbbm{1},
  L_{\mu}, g_1 W_{\mu} + L_{\mu}, B^0_{\mu} \right] \psi^{\text{c}}}, 
  \label{de}
\end{eqnarray}
where the integration over $\Sigma$ has been performed, yielding a constant
factor since the action does not depend on these fields anymore.

To determine the anomalous contribution of fermion triangles to the
three-point functions of the techni-currents, we may proceed as follows: first
one uses the expression for the techni-currents $J_L^{\mu}$ in terms of the
low-energy variables. These are obtained by considering the full lagrangian in
a generic gauge, taking the functional derivative with respect to the sources
$L_{\mu}$ {\footnote{While the currents $J_R^{\mu}$ are perfectly sensible in
the techni-theory, their expression in terms of low-energy variables would be
of little use in the present framework since the corresponding sources
$R^{\mu}$ are identified with the $\tmop{SU} \left( 2 \right)$ gauge fields
$g_1 G_1^{\mu}$.}}, and then injecting the solution to the constraints in the
standard gauge to make the field content explicit. When this is written in
terms of unitary gauge variables, we obtain {\footnote{The necessary
diagonalization for vector fields was described in {\cite{Hirn:2004ze}}.}}
\begin{eqnarray}
  J_L^{3 \nu} & = & - \sqrt{1 + \gamma^2}  \frac{M_W^2}{g_1} Z^{\nu} +
  \delta_L  \overline{\psi^{\text{c}}_L} \gamma^{\nu} \frac{\tau^3}{2}
  \psi^{\text{c}}_L + \left( 1 - \delta_R \right) 
  \overline{\psi^{\text{c}}_R} \gamma^{\nu} \frac{\tau^3}{2}
  \psi^{\text{c}}_R,  \label{dg}\\
  J_L^{\pm \nu} & = & - \frac{M_W^2}{g_1} W^{\nu \pm} + \frac{1}{\sqrt{2}}
  \delta_L  \overline{\psi^{\text{c}}_L} \gamma^{\nu} \tau^{\pm}
  \psi^{\text{c}}_L + \frac{1}{\sqrt{2}}  \left( 1 - \delta_R \right) 
  \overline{\psi^{\text{c}}_R} \gamma^{\nu} \tau^{\pm} \psi^{\text{c}}_R, 
  \label{dh}
\end{eqnarray}
where
\begin{eqnarray}
  M_W^2 & = & g_1^2  \frac{f^2}{4}, \\
  \gamma & = & \frac{g_0}{g_1} . 
\end{eqnarray}
We can then express, following {\cite{Coleman:1982yg}}, the forward matrix element of
these currents between two asymptotic fermion states described by spinors $u'$
and $u$ satisfying the massless Dirac equation, with identical momentum $p$
\begin{eqnarray}
  \left\langle u', i, p \left| J_L^{a \mu} \left( 0 \right) \right| u, j, p
  \right\rangle & = & \overline{u'} \gamma^{\mu} \mathcal{J}_{i j}^a u, 
  \label{di}
\end{eqnarray}
where $\mathcal{J}^a$ is a two by two matrix representing the charges
associated to the Noether currents~$J^{a \mu}_L$ (the indices $i, j$
distinguish between the upper and lower component of the doublet).

\DOUBLEFIGURE{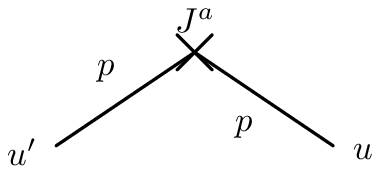}{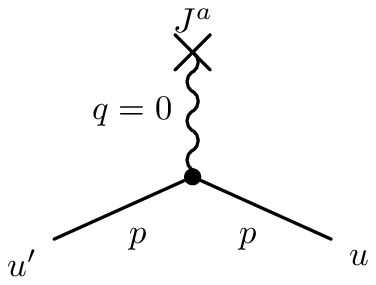}{\label{fig:direct}Direct contribution to the matrix element.}{\label{fig:indirect}Indirect contribution to the matrix element, involving the emission of a vector boson.}

To calculate this, we must take into account two contributions: one coming
from a direct contraction of the external spinors with the last two terms in
(\ref{dg}) or in (\ref{dh}), as depicted in Figure \ref{fig:direct}, and one
where a massive vector propagator is involved, as in Figure \ref{fig:indirect}.
However in this last case, the propagator is taken at $k = 0$, and this
results in a local term. The matrix element of (\ref{di}) is then given by the
sum of the two terms. In the second graph (Figure \ref{fig:indirect}), the
longitudinal polarization of the vector propagator does not contribute, as can
be seen by making use of the massless Dirac equation. Calculation of this last
diagram requires the expression for the interactions of composite fermions
with the vector fields. This is obtained by injecting the solution to the
constraints in the standard gauge into the lagrangian for composite fermions
(\ref{ar}), rewriting in the unitary gauge and then diagonalizing the vector
fields, to arrive at
\begin{eqnarray}
  \mathcal{L}_{\text{comp}} & = & \mathi \overline{\psi^{\text{c}}}
  \gamma^{\mu} \partial_{\mu} \psi^{\text{c}} + e \overline{\psi^{\text{c}}}
  \gamma^{\mu} Q \psi^{\text{c}} A_{\mu} - g_1 \frac{\gamma^2}{\sqrt{1 +
  \gamma^2}}  \overline{\psi^{\text{c}}} \gamma^{\mu} Q \psi^{\text{c}}
  Z_{\mu} \nonumber\\
  & + & g_1  \left( 1 - \delta_L  \right) \sqrt{1 + \gamma^2} 
  \overline{\psi^{\text{c}}_L} \gamma^{\mu}  \frac{\tau^3}{2}
  \psi^{\text{c}}_L Z_{\mu} + \frac{g_1}{\sqrt{2}}  \left( 1 - \delta_L
  \right)  \overline{\psi^{\text{c}}_L} \gamma^{\mu} \tau^{\pm}
  \psi^{\text{c}}_L W_{\mu}^{\pm} \nonumber\\
  & + & g_1 \delta_R  \sqrt{1 + \gamma^2}  \overline{\psi^{\text{c}}_R}
  \gamma^{\mu}  \frac{\tau^3}{2} \psi^{\text{c}}_R Z_{\mu} +
  \frac{g_1}{\sqrt{2}} \delta_R  \overline{\psi^{\text{c}}_R} \gamma^{\mu}
  \tau^{\pm} \psi^{\text{c}}_R W_{\mu}^{\pm},  \label{de1}
\end{eqnarray}
with
\begin{eqnarray}
  Q & = & \frac{\tau^3}{2} + \frac{B - L}{2},  \label{df}\\
  e & = & \frac{g_1 g_0}{\sqrt{g_1^2 + g_0^2}} . 
\end{eqnarray}
Setting $\delta_L = \delta_R = 0$ in (\ref{de1}), one may check that the same
tree-level couplings of fermions to vector fields as in the SM are recovered.
As announced before, $\delta_L$ and $\delta_R$ respectively allow for
non-universal couplings of left-handed composite fermions and for couplings of
right-handed composite fermions with $W^{\pm}$ respectively.
With these results in mind, one finds that the forward matrix elements of the type (\ref{di}) have a
vector structure and do not involve the anomalous $\delta_L$, $\delta_R$
couplings
\begin{eqnarray}
  \mathcal{J}^3 & = & \frac{1}{1 + \gamma^2}  \left( \frac{\tau^3}{2} -
  \gamma^2  \frac{B - L}{2} \right),  \label{di12}\\
  \mathcal{J}^{\pm} & = & \frac{\tau^{\pm}}{\sqrt{2}} .  \label{di13}
\end{eqnarray}
The same is true of the matrix elements of the electromagnetic current. Thus,
triangle diagrams with composite fermions running around the loop do not
generate any anomalous contribution to the three-point functions of the
techni-currents and electromagnetic currents. There are no other possible
anomalous contributions to these Green's functions in unitary gauge since the
Wess-Zumino term $\hat{S}_{\tmop{WZ}}$ occurring in the right-hand side of
(\ref{de}) vanishes due to the boundary condition (\ref{bg}). This result
agrees with that of section~\ref{sec: obstruction}, which stated that the
Green's functions of the techni-currents must be anomaly-free, otherwise we
would not have been able to absorb the Goldstone bosons when defining the
unitary gauge in the first place. The effects of $\delta_L$ and $\delta_R$
will then only be visible for~$q^2 \neq 0$ in figure \ref{fig:indirect}.

On the other hand, we still have to investigate the contribution of triangular
diagrams to the three-point functions of the gauge bosons: anomalies were
first encountered in this context in the literature. Expressing the couplings
of fermions to gauge bosons in unitary gauge as given by (\ref{de1}) in terms
of the following building blocks
\begin{eqnarray}
  j_{\mu}^0 & = & \frac{1}{2}  \overline{\psi^{\text{c}}} \gamma_{\mu}
  \psi^{\text{c}},  \label{dj}\\
  j_{\mu}^a & = & \overline{\psi^{\text{c}}} \gamma_{\mu}  \frac{\tau^a}{2}
  \psi^{\text{c}},  \label{dk}\\
  j_{\mu}^{5, a} & = & \overline{\psi^{\text{c}}} \gamma_{\mu} \gamma_5 
  \frac{\tau^a}{2} \psi^{\text{c}},  \label{dl}
\end{eqnarray}
one can give the expression of the triangle contribution to an effective
triple gauge boson interaction in terms of one single function $T_{\mu \nu
\rho}$ (see {\cite{Coleman:1982yg}}) containing the full information about the
triangle
\begin{eqnarray}
  &  & ( 2 \mathpi )^4 \delta^{\left( 4 \right)} \left( k_1 + k_2 + k_3
  \right) T_{\mu \nu \rho} \left( k_1, k_2, k_3 \right) \nonumber\\
  & = & - \int \mathd x \mathd y \mathd z \mathe^{- \mathi \left( k_1 \cdot x
  + k_2 \cdot y + k_3 \cdot z \right)}  \left\langle 0 \left| Tj^0_{\mu}
  \left( x \right) j_{\nu}^3 \left( y \right) j_{\rho}^{5, 3} \left( z \right)
  \right| 0 \right\rangle .  \label{dj1}
\end{eqnarray}
Defining
\begin{eqnarray}
  \mathe^{\mathi W \left[ W_{\mu}^{\pm}, Z_{\mu}, A_{\mu} \right]} & = & \int
  \mathd \left[ \psi^{\text{c}} \right] \mathe^{\mathi \int \mathd
  x\mathcal{L}_{\tmop{comp}. \tmop{ferm}.} \left[ \psi^{\text{c}},
  W_{\mu}^{\pm}, Z_{\mu}, A_{\mu} \right] } , 
\end{eqnarray}
we get in momentum space the following trilinear terms
\begin{eqnarray}
  W \left[ W_{\mu}^{\pm}, Z_{\mu}, A_{\mu} \right] & = & \int \frac{\mathd
  k_2}{\left( 2 \mathpi \right)^4}  \frac{\mathd k_3}{\left( 2 \mathpi
  \right)^4} g_1^3 T_{\mu \nu \rho} \left( - k_2 - k_3, k_2, k_3 \right)
  \nonumber\\
  & \times & \left\{ D^{A W W} A^{\mu} \left( - k_2 - k_3 \right)  \left(
  W^{+ \nu} \left( k_2 \right) W^{- \rho} \left( k_3 \right) + W^{- \nu}
  \left( k_2 \right) W^{+ \rho} \left( k_3 \right) \right) \right. \nonumber\\
  & + & D^{Z W W} Z^{\mu} \left( - k_2 - k_3 \right)  \left( W^{+ \nu} \left(
  k_2 \right) W^{- \rho} \left( k_3 \right) + W^{- \nu} \left( k_2 \right)
  W^{+ \rho} \left( k_3 \right) \right) \nonumber\\
  & + & D^{Z Z Z} Z^{\mu} \left( - k_2 - k_3 \right) Z^{\nu} \left( k_2
  \right) Z^{\rho} \left( k_3 \right) \nonumber\\
  & + & D^{A Z Z} A^{\mu} \left( - k_2 - k_3 \right) Z^{\nu} \left( k_2
  \right) Z^{\rho} \left( k_3 \right) \nonumber\\
  & + & D^{Z A Z} Z^{\mu} \left( - k_2 - k_3 \right) A^{\nu} \left( k_2
  \right) Z^{\rho} \left( k_3 \right) \nonumber\\
  & + & D^{A A Z} A^{\mu} \left( - k_2 - k_3 \right) A^{\nu} \left( k_2
  \right) Z^{\rho} \left( k_3 \right) \nonumber\\
  & + & \left. \text{quadratic terms and higher powers} \right\}, 
  \label{dk1}
\end{eqnarray}
corresponding to the triangular diagrams. In (\ref{dk1}), the $D$ constants
are determined as linear combinations of only two constants
\begin{eqnarray}
  D^{A W W} & = & \frac{\gamma}{\sqrt{1 + \gamma^2}}  \frac{B - L}{4}  \left(
  \delta_R^2 - \left( 1 - \delta_L \right)^2 \right),  \label{dl1}\\
  D^{A A Z} & = & \frac{\gamma^2}{\sqrt{1 + \gamma^2}}  \frac{B - L}{2} 
  \left( \delta_R + \delta_L - 1 \right),  \label{dm}
\end{eqnarray}
by the following relations
\begin{eqnarray}
  D^{Z W W} & = & - \gamma D^{A W W},  \label{dn}\\
  D^{Z A Z} & = & - \gamma D^{A A Z},  \label{do}\\
  D^{A Z Z} & = & \left( 1 + \gamma^2 \right) D^{A W W} - \gamma D^{A A Z}, 
  \label{dp}\\
  D^{Z Z Z} & = & - \gamma \left( \left( 1 + \gamma^2 \right) D^{A W W} -
  \gamma D^{A A Z} \right) .  \label{dq}
\end{eqnarray}
Thus we find that these loops do in particular generate trilinear interactions
between neutral electroweak vector bosons. The implications of anomalies and
of the parameters~$\delta_L, \delta_R$ on these interactions may then be
studied following {\cite{Renard:1982es,Barroso:1985re,Gounaris:1999kf,Gounaris:2000tb}}. Let us just point out that anomalous contributions from
each doublet would vanish if {\footnote{There is no contradiction with the
fact that elementary fermions have $\delta_L = \delta_R = 0$ at this order:
for elementary fermions, the absence of anomalous contributions to three-point
functions of the electroweak vector bosons follows from the condition on the
trace of $B - L$.}}
\begin{eqnarray}
  \delta_L + \delta_R & = & 1 . 
\end{eqnarray}

Note that renormalizability order-by-order in the sense of Weinberg
{\cite{Weinberg:1979kz}} is not lost: on the contrary, we have seen that the
three-point function of the techni-currents obey the appropriate
transformation properties as deduced from the techni-theory, and which serve
as a guide for the construction of the effective lagrangian. On the other
hand, we find that there are in general anomalous contributions to three-point
functions of electroweak vector bosons. It should be stressed that there is no
conflict here: the symmetry of the techni-theory on which the LEET hinges must
be anomaly-free, and the generating functional must reflect this. On the other
hand, the symmetries whose currents are coupled to the elementary gauge bosons
may be anomalous. Since the  effective theory framework is not based on the
requirement of renormalizability in the usual sense, this does not endanger
the consistency of the theory.

\section{Conclusion} \label{s:concl}

In this paper, we have investigated the minimal low-energy effective theory
description of EWSB, where renormalization proceeds order-by-order in the
momentum expansion instead of the usual renormalization order-by-order in
powers of the coupling constants. In particular, we have analyzed the relation
implied by the anomaly-matching between the effective theory operating with
low-energy degrees of freedom (Goldstone bosons and, possibly, light composite
fermions) and the fundamental `techni-theory', representing the unknown
ultra-violet completion of the LEET.

We have derived a necessary condition, which the techni-theory has to satisfy
in order to allow all Goldstone modes to be removed from the spectrum by the
Higgs mechanism: symmetries of the techni-theory under which the low-energy
degrees of freedom are charged must be anomaly-free, that is $k_{\tmop{TT}} =
0$ in the notation of section \ref{s:anomalies}. This conclusion is true
independently of the presence of composite fermions: the anomaly-matching 
allowed us to phrase the question directly in terms of the underlying
techni-theory.
Technicolor models are usually constructed to be anomaly-free~\cite{Farhi:1981xs}. The relevant question is then how the restriction $k_{\tmop{TT}}=0$ modifies the spectrum of low-lying bound states as compared to QCD in which $k_{\tmop{TT}}=N_c/3$. Beyond such models, translating the restriction~$k_{\tmop{TT}} = 0$ in terms of the fundamental variables of a generic techni-theory is far from obvious: to achieve this, one would have to go beyond the effective theory.

Concerning the anomaly-matching, we have extended the construction of the
Wess-Zumino effective lagrangian to the case where the low-energy sector
contains composite fermions in addition to Goldstone bosons. This turns out to
be possible whatever the values of the $\mathrm{U} \left( 1 \right)_{B - L}$
charges of the composite fermions, and for any values of their non-standard
couplings $\delta_L, \delta_R$. Consequently, there are no restrictions on
these couplings. The possibility of having an (incomplete) fourth generation
of fermions, which would be composite is thus theoretically open.

Within the spurion formalism developed in {\cite{Hirn:2004ze}}, the
distinction between composite and elementary fermions is unambiguous: since
the composite and elementary sectors of the theory are coupled only via
spurions, the mixing between composite and elementary fermions is an effect of
higher orders in the spurion expansion. Qualitatively, the exclusive
properties of composite fermions should be: i) the non-standard couplings
$\delta_L$, $\delta_R$ describing a violation of universality of left-handed
couplings to electroweak vector bosons, as well as right-handed couplings to
$W^{\pm}$, ii) a relatively large mean mass $m$ of the doublet and a splitting
within the doublet $\Delta m \ll m$, iii) no restriction on $B - L$, that is, a
possibility to have incomplete composite fermion generations. These properties
practically exclude the interpretation of known fermion doublets as composite
in the above sense. We have also shown in section \ref{s:triangles} that composite
fermions would indirectly manifest themselves in the triple boson vertices
induced by anomalous triangular graphs. Only composite fermions that have
$\delta_L + \delta_R \neq 1$ can contribute to such graphs.

Elementary fermions necessarily have $\delta_L = \delta_R = 0$ at the
leading-order of the spurion expansion. They do not participate in the
anomaly-matching, being exterior to the composite sector. Nevertheless, as
shown in section \ref{s:U-gauge}, one recovers the usual condition that the
trace of $B - L$ over elementary fermions vanishes. The reason for this is not
the usual requirement of renormalizability, but rather the assumption that all
Goldstone modes are removed from the spectrum by the Higgs mechanism. The
whole effective theory can then be consistently formulated in the unitary
gauge.

{\section*{Acknowledgments}}

We are grateful to Marc Knecht for valuable discussions. We also profited from
stimulating questions and comments from Laurent Lellouch.
We express our thanks to the referee for helping us in straightening out our argument  concerning technicolor models as found in the literature.

This work was supported in part by the European Community EURIDICE network
under contract HPRN-CT-2002-00311.

\bibliographystyle{JHEP}
\bibliography{biblio.bib}

\providecommand{\href}[2]{#2}\begingroup\raggedright\begin{thebibliography}{10}

\bibitem{Weinberg:1979kz}
S.~Weinberg, {\it Phenomenological lagrangians},  {\em Physica} {\bf A96}
  (1979) 327.

\bibitem{Gasser:1984yg}
J.~Gasser and H.~Leutwyler, {\it Chiral {P}erturbation {T}heory to one loop},
  {\em Ann. Phys.} {\bf 158} (1984) 142.

\bibitem{Gasser:1985gg}
J.~Gasser and H.~Leutwyler, {\it Chiral {P}erturbation {T}heory: {e}xpansions
  in the mass of the strange quark},  {\em Nucl. Phys.} {\bf B250} (1985) 465.

\bibitem{Georgi:1985kw}
H.~Georgi, {\em Weak interactions and modern particle theory}.
\newblock Benjamin/Cummings Publishing, 1984.

\bibitem{Dobado:1991zh}
A.~Dobado, D.~Espriu, and M.~J. Herrero, {\it Chiral lagrangians as a tool to
  probe the symmetry breaking sector of the {SM} at {LEP}},  {\em Phys. Lett.}
  {\bf B255} (1991) 405.

\bibitem{Nyffeler:1999hp}
A.~Nyffeler, {\it The electroweak chiral lagrangian revisited},
  \href{http://xxx.lanl.gov/abs/hep-ph/9912472}{{\tt hep-ph/9912472}}.

\bibitem{Susskind:1979ms}
L.~Susskind, {\it Dynamics of spontaneous symmetry breaking in the
  {W}einberg-{S}alam theory},  {\em Phys. Rev.} {\bf D20} (1979) 2619.

\bibitem{Holdom:1990tc}
B.~Holdom and J.~Terning, {\it Large corrections to electroweak parameters in
  technicolor theories},  {\em Phys. Lett.} {\bf B247} (1990) 88.

\bibitem{GOlden:1991ig}
M.~Golden and L.~Randall, {\it Radiative corrections to electroweak parameters
  in technicolor theories},  {\em Nucl. Phys.} {\bf B361} (1991) 3.

\bibitem{Peskin:1992sw}
M.~E. Peskin and T.~Takeuchi, {\it Estimation of oblique electroweak
  corrections},  {\em Phys. Rev.} {\bf D46} (1992) 381.

\bibitem{Hagiwara:2002fs}
{\bf Particle Data Group} Collaboration, K.~Hagiwara {\em et.~al.}, {\it Review
  of particle physics},  {\em Phys. Rev.} {\bf D66} (2002) 010001.

\bibitem{Hill:2002ap}
C.~T. Hill and E.~H. Simmons, {\it Strong dynamics and electroweak symmetry
  breaking},  {\em Phys. Rept.} {\bf 381} (2003) 235--402,
  [\href{http://xxx.lanl.gov/abs/hep-ph/0203079}{{\tt hep-ph/0203079}}].

\bibitem{Nyffeler:1999ap}
A.~Nyffeler and A.~Schenk, {\it The electroweak chiral lagrangian reanalyzed},
  {\em Phys. Rev.} {\bf D62} (2000) 113006,
  [\href{http://xxx.lanl.gov/abs/hep-ph/9907294}{{\tt hep-ph/9907294}}].

\bibitem{Holdom:1981rm}
B.~Holdom, {\it Raising the sideways scale},  {\em Phys. Rev.} {\bf D24} (1981)
  1441.

\bibitem{Holdom:1985sk}
B.~Holdom, {\it Techniodor},  {\em Phys. Lett.} {\bf B150} (1985) 301.

\bibitem{Yamawaki:1986zg}
K.~Yamawaki, M.~Bando, and K.-I. Matumoto, {\it Scale-invariant hypercolor
  model and a idilaton},  {\em Phys. Rev. Lett.} {\bf 56} (1986) 1335.

\bibitem{Appelquist:1986an}
T.~W. Appelquist, D.~Karabali, and L.~C.~R. Wijewardhana, {\it Chiral
  hierarchies and the flavor changing neutral current problem in technicolor},
  {\em Phys. Rev. Lett.} {\bf 57} (1986) 957.

\bibitem{Appelquist:1987tr}
T.~Appelquist and L.~C.~R. Wijewardhana, {\it Chiral hierarchies and chiral
  perturbations in technicolor},  {\em Phys. Rev.} {\bf D35} (1987) 774.

\bibitem{Appelquist:1980vg}
T.~Appelquist and C.~W. Bernard, {\it Strongly interacting {H}iggs boson},
  {\em Phys. Rev.} {\bf D22} (1980) 200.

\bibitem{Longhitano:1980iz}
A.~C. Longhitano, {\it Heavy {H}iggs bosons in the {W}einberg-{S}alam model},
  {\em Phys. Rev.} {\bf D22} (1980) 1166.

\bibitem{Longhitano:1981tm}
A.~C. Longhitano, {\it Low-energy impact of a heavy {H}iggs boson sector},
  {\em Nucl. Phys.} {\bf B188} (1981) 118.

\bibitem{Csaki:2003zu}
C.~Cs{\'a}ki, C.~Grojean, L.~Pilo, and J.~Terning, {\it Towards a realistic
  model of {H}iggsless electroweak symmetry breaking},
  \href{http://xxx.lanl.gov/abs/hep-ph/0308038}{{\tt hep-ph/0308038}}.

\bibitem{Barbieri:2003pr}
R.~Barbieri, A.~Pomarol, and R.~Rattazzi, {\it Weakly coupled {H}iggsless
  theories and precision electroweak tests},
  \href{http://xxx.lanl.gov/abs/hep-ph/0310285}{{\tt hep-ph/0310285}}.

\bibitem{Cacciapaglia:2004jz}
G.~Cacciapaglia, C.~Csaki, C.~Grojean, and J.~Terning, {\it Oblique corrections
  from Higgsless models in warped space},
  \href{http://xxx.lanl.gov/abs/hep-ph/0401160}{{\tt hep-ph/0401160}}.

\bibitem{Hirn:2004ze}
J.~Hirn and J.~Stern, {\it The role of spurions in {H}iggsless electroweak
  effective theories},  {\it Eur. Phys. J.} {\bf C 34} (2004) 447, \href{http://xxx.lanl.gov/abs/hep-ph/0401032}{{\tt  hep-ph/0401032}}.

\bibitem{'tHooft:1979bh}
G.~'t~Hooft, {\it Naturalness, chiral symmetry, and spontaneous chiral symmetry
  breaking},  in G.~'t~Hooft {\em  et.~al.}, eds., {\em Recent developments in gauge theories. Proceedings of the {NATO} Advanced Study Institute, Carg{\`e}se, 1979}. Plenum Press, 1979.
\newblock .

\bibitem{Coleman:1982yg}
S.~R. Coleman and B.~Grossman, {\it 't {H}ooft's consistency condition as a
  consequence of analyticity and unitarity},  {\em Nucl. Phys.} {\bf B203}
  (1982) 205.

\bibitem{Wess:1971yu}
J.~Wess and B.~Zumino, {\it Consequences of anomalous {W}ard identities},  {\em
  Phys. Lett.} {\bf B37} (1971) 95.

\bibitem{Leutwyler:1994iq}
H.~Leutwyler, {\it On the foundations of {C}hiral {P}erturbation {T}heory},
  {\em Ann. Phys.} {\bf 235} (1994) 165,
  [\href{http://xxx.lanl.gov/abs/hep-ph/9311274}{{\tt hep-ph/9311274}}].

\bibitem{Wudka:1994ny}
J.~Wudka, {\it Electroweak effective lagrangians},  {\em Int. J. Mod. Phys.}
  {\bf A9} (1994) 2301, [\href{http://xxx.lanl.gov/abs/hep-ph/9406205}{{\tt
  hep-ph/9406205}}].

\bibitem{Appelquist:1985rr}
T.~Appelquist, M.~J. Bowick, E.~Cohler, and A.~I. Hauser, {\it The breaking of
  isospin symmetry in theories with a dynamical {H}iggs mechanism},  {\em Phys.
  Rev.} {\bf D31} (1985) 1676.

\bibitem{Peccei:1990kr}
R.~D. Peccei and X.~Zhang, {\it Dynamical symmetry breaking and universality
  breakdown},  {\em Nucl. Phys.} {\bf B337} (1990) 269.

\bibitem{Urech:1995hd}
R.~Urech, {\it Virtual photons in {C}hiral {P}erturbation {T}heory},  {\em
  Nucl. Phys.} {\bf B433} (1995) 234,
  [\href{http://xxx.lanl.gov/abs/hep-ph/9405341}{{\tt hep-ph/9405341}}].

\bibitem{Weinberg:1996kr}
S.~Weinberg, {\em The quantum theory of fields, Vol. 2: Modern applications}.
\newblock Cambridge Univ. Press, 1996.

\bibitem{Adler:1969gk}
S.~L. Adler, {\it Axial-vector vertex in spinor electrodynamics},  {\em Phys.
  Rev.} {\bf 177} (1969) 2426.

\bibitem{Bardeen:1969md}
W.~A. Bardeen, {\it Anomalous {W}ard identities in spinor field theories},
  {\em Phys. Rev.} {\bf 184} (1969) 1848.

\bibitem{Adler:1969er}
S.~L. Adler and W.~A. Bardeen, {\it Absence of higher-order corrections in the
  anomalous axial-vector divergence equation},  {\em Phys. Rev.} {\bf 182}
  (1969) 1517.

\bibitem{Fujikawa:1979ay}
K.~Fujikawa, {\it Path-integral measure for gauge invariant fermion theories},
  {\em Phys. Rev. Lett.} {\bf 42} (1979) 1195.

\bibitem{Fujikawa:1980eg}
K.~Fujikawa, {\it Path integral for gauge theories with fermions},  {\em Phys.
  Rev.} {\bf D21} (1980) 2848.

\bibitem{Balachandran:1982cs}
A.~P. Balachandran, G.~Marmo, V.~P. Nair, and C.~G. Trahern, {\it
  Nonperturbative proof of the nonabelian anomalies},  {\em Phys. Rev.} {\bf
  D25} (1982) 2713.

\bibitem{Tsutsui:1989qr}
I.~Tsutsui, {\it Origin of anomalies in the path integral formalism},  {\em
  Phys. Rev.} {\bf D40} (1989) 3543.

\bibitem{Kaiser:2000ck}
R.~Kaiser, {\it Anomalies and {WZW}-term of two-flavour {QCD}},  {\em Phys.
  Rev.} {\bf D63} (2001) 076010,
  [\href{http://xxx.lanl.gov/abs/hep-ph/0011377}{{\tt hep-ph/0011377}}].

\bibitem{Banks:1980sc}
T.~Banks, S.~Yankielowicz, and A.~Schwimmer, {\it Anomaly constraints in chiral
  gauge theories},  {\em Phys. Lett.} {\bf B96} (1980) 67.

\bibitem{Frishman:1981dq}
Y.~Frishman, A.~Schwimmer, T.~Banks, and S.~Yankielowicz, {\it The axial
  anomaly and the bound state spectrum in confining theories},  {\em Nucl.
  Phys.} {\bf B177} (1981) 157.

\bibitem{Alvarez-Gaume:1985yb}
L.~Alvarez-Gaum{\'e} and P.~Ginsparg, {\it Geometry anomalies},  {\em Nucl.
  Phys.} {\bf B262} (1985) 439.

\bibitem{Gross:1972pv}
D.~J. Gross and R.~Jackiw, {\it Effect of anomalies on quasirenormalizable
  theories},  {\em Phys. Rev.} {\bf D6} (1972) 477.

\bibitem{Grosse-Knetter:1993nn}
C.~Grosse-Knetter and R.~K{\"{o}}gerler, {\it Unitary gauge,
  {S}t{\"{u}}ckelberg formalism and gauge invariant models for effective
  lagrangians},  {\em Phys. Rev.} {\bf D48} (1993) 2865,
  [\href{http://xxx.lanl.gov/abs/hep-ph/9212268}{{\tt hep-ph/9212268}}].

\bibitem{Farhi:1981xs}
E.~Farhi and L.~Susskind, {\it Technicolor},  {\em Phys. Rept.} {\bf 74} (1981)
  277.

\bibitem{Bouchiat:1972iq}
C.~Bouchiat, J.~Iliopoulos, and P.~Meyer, {\it An anomaly free version of
  {W}einberg's model},  {\em Phys. Lett.} {\bf B38} (1972) 519.

\bibitem{KorthalsAltes:1972aq}
C.~P. Korthals~Altes and M.~Perrottet, {\it Anomalous {W}ard identities,
  gauge-variance and appearance of ghosts in {H}iggs-{K}ibble type theories},
  {\em Phys. Lett.} {\bf B39} (1972) 546.

\bibitem{Georgi:1972bb}
H.~Georgi and S.~L. Glashow, {\it Gauge theories without anomalies},  {\em
  Phys. Rev.} {\bf D6} (1972) 429.

\bibitem{Renard:1982es}
F.~M. Renard, {\it Tests of neutral gauge boson self-couplings with {$e^+ e^-
  \to \gamma Z$}},  {\em Nucl. Phys.} {\bf B196} (1982) 93.

\bibitem{Barroso:1985re}
A.~Barroso, F.~Boudjema, J.~Cole, and N.~Dombey, {\it Electromagnetic
  properties of the {$Z$} boson. 1},  {\em Z. Phys.} {\bf C28} (1985) 149.

\bibitem{Gounaris:1999kf}
G.~J. Gounaris, J.~Layssac, and F.~M. Renard, {\it Signatures of the anomalous
  {$Z \gamma$} and {$Z Z$} production at the lepton and hadron colliders},
  {\em Phys. Rev.} {\bf D61} (2000) 073013,
  [\href{http://xxx.lanl.gov/abs/hep-ph/9910395}{{\tt hep-ph/9910395}}].

\bibitem{Gounaris:2000tb}
G.~J. Gounaris, J.~Layssac, and F.~M. Renard, {\it New and standard physics
  contributions to anomalous {$Z$} and {$\gamma$} self-couplings},  {\em Phys.
  Rev.} {\bf D62} (2000) 073013,
  [\href{http://xxx.lanl.gov/abs/hep-ph/0003143}{{\tt hep-ph/0003143}}].

\end{thebibliography}\endgroup

\end{document}